\documentclass[traditabstract]{aa} 
    
\usepackage{graphicx}
 \usepackage{longtable,lscape}
\usepackage{natbib}
\usepackage{color}
\usepackage{pbox}
\usepackage{amssymb}
\usepackage{amsmath}
\usepackage[table]{xcolor}
\usepackage{listings}

\newcommand{\kms}{{\rm km}\,{\rm s}^{-1}}

\newcommand{\msun}{{M}_\odot}

\usepackage{txfonts}
\begin{document}
   \title{Binary-corrected velocity dispersions from single- and multi-epoch radial velocities: massive stars in R136 as a test case}

   \author{M. Cottaar\inst{1} \and V. H\'{e}nault-Brunet\inst{2, 3}}

   \institute{Institute for Astronomy, ETH Zurich, Wolfgang-Pauli-Strasse 27, 8093 Zurich, Switzerland; \email{MCottaar@phys.ethz.ch}    
   \and
   Department of Physics, Faculty of Engineering and Physical Sciences, University of Surrey, Guildford, GU2 7XH, UK; \email{v.henault-brunet@surrey.ac.uk}      
   \and
   Scottish Universities Physics Alliance (SUPA), Institute for Astronomy, University of Edinburgh, Blackford Hill, Edinburgh, EH9 3HJ, UK 
      }

   \date{Received ; accepted }

\abstract
{Orbital motions from binary stars can broaden the observed line-of-sight velocity distribution of a stellar system, artificially inflating the measured line-of-sight velocity dispersion, which can in turn lead to erroneous conclusions about the dynamical state of the system. Cottaar et al. (2012b) proposed a maximum likelihood procedure to recover the intrinsic velocity dispersion of a resolved star cluster from a single epoch of radial velocity data of individual stars, which they achieved by simultaneously fitting the intrinsic velocity distribution of the single stars and the centres of mass of the binaries along with the velocity shifts caused by binary orbital motions. Assuming well-characterized binary properties, they showed that this procedure can accurately reproduce intrinsic velocity dispersions below $1\ \kms$ for solar-type stars. Here we investigate the systematic offsets induced in cases where the binary properties are uncertain, and we show how two epochs of radial velocity data with an appropriate baseline can help to mitigate these systematic effects. We first test the method above using Monte Carlo simulations, taking into account the large uncertainties in the binary properties of OB stars. We then apply it to radial velocity data in the young massive cluster R136, an example for which the intrinsic velocity dispersion of O-type stars is known from an intensive multi-epoch approach. For typical velocity dispersions of young massive clusters ($\gtrsim 4 \ \kms$) and with a single epoch of data, we demonstrate that the method can just about distinguish between a cluster in virial equilibrium and an unbound cluster. This is due to the higher spectroscopic binary fraction and more loosely constrained distributions of orbital parameters of OB stars compared to solar-type stars. By extending the maximum likelihood method to multi-epoch data, we show that the accuracy on the fitted velocity dispersion can be improved to a few percent by using only two epochs of radial velocities. This procedure offers a promising method of accurately measuring the intrinsic stellar velocity dispersion in other systems for which the binary properties are not well constrained, for example young clusters and associations whose luminosity is dominated by OB stars.}


\authorrunning{Cottaar \& H\'{e}nault-Brunet}\titlerunning{Binary-corrected velocity dispersion from single- and multi-epoch radial velocities}

   \maketitle
%

\section{Introduction}

An accurate measurement of the intrinsic velocity dispersion of a star cluster is required to estimate its virial state, which determines whether a cluster is currently bound or whether it is dissolving into the field. Recent observational studies of resolved young massive clusters suggested that the majority have velocity dispersions of a few $\kms$ and seem to be in virial equilibrium. These studies are based both on proper motions \citep[e.g.][]{rochau2010, clarkson2012} and multi-epoch radial velocities \citep[e.g.][]{cottaar2012a, VHB2012}. The luminosity of these young massive clusters is dominated by massive OB stars, a large fraction of which are spectroscopic binaries \citep[e.g.][]{KK2012, Sana2012_science, Sana_2013}, so the radial velocity studies have been designed to include multiple epochs. With this multi-epoch strategy, the significant effect of spectroscopic binaries on the observed radial velocity distribution \citep[e.g.][]{gieles2010} can be reduced by identifying as many of these systems as possible and removing them from the sample. Repeated radial velocity measurements have also been used to correct for the influence of binary stars in other types of systems, for example in dwarf galaxies \citep[e.g.][]{simon2011, koposov2011}.

Alternatively, one can use a maximum likelihood procedure to recover the intrinsic velocity dispersion of stellar system from a single epoch of radial velocity data despite the orbital motions of spectroscopic binaries by simultaneously fitting the intrinsic velocity distribution of the single stars and centres of mass of the binaries along with the radial velocities caused by orbital motions \citep{Odenkirchen02, Kleyna02, Martinez11, cottaar2012b}. This technique allows one to study the dynamical state of clusters having a low velocity dispersion without the need for multi-epoch observations to identify the spectroscopic binaries. Even when multi-epoch data is available, the same technique can still be used (with minor modifications, see Sec \ref{sec:extension}) to correct for the orbital motions of undetected binaries.

\citet{cottaar2012b} showed that this method could successfully reproduce the intrinsic velocity dispersion of 0.5~$\kms$ of the old open cluster NGC\,188 \citep{Geller2008, Geller2009, GellerMathieu2011} using a single epoch of radial velocity data and adopting the observed orbital parameter distributions (i.e. period, mass ratio, and eccentricity) for solar-type field stars \citep{Raghavan10, Reggiani13}. It is not obvious that the same method can be applied to successfully recover the intrinsic velocity dispersion of young massive clusters due to the fact that (1) the OB stars dominating the light of these systems have a high spectroscopic binary fraction \citep[e.g.][and references therein]{Sana2012_science}, (2) the distributions of orbital parameters of massive binaries are relatively loosely constrained (compared to solar-type stars), (3) higher masses imply larger contaminations to the velocity dispersion as the orbital velocity scales with the mass of the primary as $v_{\rm orb} \propto M^{1/3}$, (4) the period distribution of massive binaries generally appears skewed towards shorter periods (see Section \ref{sec:binaries}), which also implies larger contaminations, and (5) the accuracy of radial velocity measurements is generally lower for OB stars (typically a few $\kms$ or more) due to rotational broadening of their spectral lines. Dedicated tests of the method using tailored Monte Carlo experiments are therefore needed. It is also desirable to perform tests on a young massive cluster for which the intrinsic velocity dispersion was measured from an intensive dataset of multi-epoch radial velocities (i.e. after selecting out spectroscopic binaries).

In this paper, we take advantage of the unique dataset provided by the VLT-FLAMES Tarantula Survey \citep[VFTS; ][]{VFTS_I}. From multi-epoch spectroscopic data of massive stars in the 30 Doradus region of the Large Magellanic Cloud, \citet{VHB2012} measured the intrinsic velocity dispersion of the young massive cluster R136 ($M\sim10^{5}$~$\msun$ - e.g. \citealt{2009ApJ...707.1347A}; age $\sim2$~Myr - e.g. \citealt{Crowther2010}, \citealt{koterheap}), corrected for the orbital motions of binaries. \citet{Sana_2013} studied the spectroscopic binary fraction and distributions of orbital parameters (period and mass ratio) of the O-type stars in the broader 30 Doradus region. Using the radial velocities measured in these previous studies, we apply the procedure of \citet{cottaar2012b} and check if the correct velocity dispersion can be recovered from a single epoch of radial velocities. In addition, we test the systematic and statistical uncertainties in the procedure through Monte Carlo simulations for both single-epoch and multi-epoch observations.

In section~\ref{method+bin_prop}, we briefly outline the maximum likelihood method presented by \citet{cottaar2012b}, introduce an extension of this procedure applicable to multi-epoch data, and discuss available constraints on the binary properties of OB stars. We describe Monte Carlo simulations used to test the procedure for single- and two-epoch datasets in section~\ref{sec:monte_carlo}. We then present the R136 dataset and the results obtained when applying the maximum likelihood method to these data in section~\ref{R136}. We discuss the implications of these results in section~\ref{discussion}, and finally present our conclusions in section~\ref{conclusion}.

\section{Method \label{method+bin_prop}}

\subsection{Outline of the maximum likelihood method \label{sec:method}}
The details of the maximum likelihood procedure used to fit an observed radial velocity distribution can be found in \citet{cottaar2012b}. We provide with the current paper the necessary python code\footnote{https://github.com/MichielCottaar/velbin} to calculate the likelihood to reproduce the observed radial velocities given an intrinsic velocity distribution, measurement uncertainties, and a set of binary orbital parameter distributions. For a given set of orbital parameter distributions, we vary the cluster's binary fraction, mean radial velocity, and intrinsic line-of-sight velocity dispersion to maximize this likelihood to reproduce the data. 

The observed radial velocity of a cluster member is the sum of three components: the radial velocity of the centre of mass, the radial velocity offset due to the orbital motion of the binary, and the measurement uncertainty. The likelihood function for the observed radial velocity is therefore given by the convolution of the probability density functions (pdf's) of these three components. Here we summarize a method to compute this convolution.

The convolution of the intrinsic radial velocity distribution of the cluster and the measurement uncertainties, assuming they are Gaussian, yields:
\begin{equation}
{\rm pdf}_{\rm single}(v_i, \sigma_i) = \frac{1}{\sqrt{2 \pi (\sigma_v^2 + \sigma_i^2)}} \exp{\left(-\frac{(v_i - \mu)^2}{2 (\sigma_v^2 + \sigma_i^2)}\right)}, \label{eq:pdf_single}
\end{equation}
where $\mu$ and $\sigma_v$ are the mean radial velocity and intrinsic line-of-sight velocity dispersion of the cluster and $v_i$ and $\sigma_i$ are the observed radial velocity and corresponding uncertainty for star $i$. For a single star without binary orbital motion this pdf directly gives the likelihood of reproducing the observed radial velocity.

For spectroscopic binaries we convolve the above pdf with the pdf of radial velocity offsets due to binary orbital motions. As no analytic form of the latter pdf is available, this convolution is done numerically. First, we generate a large number ($N$) of random binaries from the adopted orbital period, mass ratio, and eccentricity distributions and calculate for each one the size of the three-dimensional velocity vector $v_k$. As these velocities are randomly orientated with respect to the line of sight, the projected velocity offset will be a random value between $-v_k$ and $+v_k$, which yields for the pdf of velocity offsets projected along the line of sight:
\begin{equation}
{\rm pdf}_{\rm offset}(v_{\rm bin}) = \frac{1}{N} \sum_k 
 \begin{cases} 
   0& \mbox{if } v_k < |v_{\rm bin}| \\ 
   \frac{1}{2 v_k} & \mbox{if } v_k > |v_{\rm bin}|,
\end{cases} \label{eq:pdf_offset}
\end{equation}
where ${\rm pdf}_{\rm offset}(v_{\rm bin})$ gives the probability distribution of the line-of-sight velocity offset $v_{\rm bin}$ due to binary orbital motions.

Equation \ref{eq:pdf_offset} can be computed for a single primary stellar mass $M_{\rm prim}$ (e.g. 1 $\msun$) and can then be easily extended to other primary stellar masses by using the fact that the orbital velocity increases proportionally to $M_{\rm prim}^{1/3}$ for a fixed period, mass ratio, and eccentricity distribution. If known, the primary mass of each observed star can be taken into account by multiplying (for each star) the velocity grid on which ${\rm pdf}_{\rm offset}(v_{\rm bin})$ was computed by $(M_{\rm prim}/M_{\odot})^{1/3}$ and dividing the ${\rm pdf}_{\rm offset}$ by the same amount to conserve the normalization \citep[see][]{cottaar2012b}. We will assume a stellar mass of 30 $M_{\odot}$ for the O-stars studied in this paper.

The pdf for the observed radial velocity distribution from binary systems can then be computed by numerically convolving the pdf for the measurement uncertainty and centre-of-mass radial velocity (equation \ref{eq:pdf_single}) and the pdf for the radial velocity offset due to the binary orbital motion (equation \ref{eq:pdf_offset}, potentially corrected for primary mass):
\begin{equation}
{\rm pdf}_{\rm binary}(v_i, \sigma_i) = \int {\rm pdf}_{\rm single}(v', \sigma_i) {\rm pdf}_{\rm offset}(v_i - v') dv', \label{eq:pdf_bin}
\end{equation}
where $v'$ is the sum of the radial velocity of the centre of mass of the binary and the contribution from the measurement uncertainty, and $v_i - v'$ is the radial velocity offset due to the binary orbital motion. In the provided code equation~\ref{eq:pdf_offset} is pre-computed on a dense grid, which allows for the repeated rapid computation of the integral in equation~\ref{eq:pdf_bin}, as long as the distributions of binary orbital parameters are not varied during the fit (which would require ${\rm pdf}_{\rm offset}$ to be recomputed).

\begin{figure}[!b]
  \begin{center}
    \includegraphics[width=.5\textwidth]{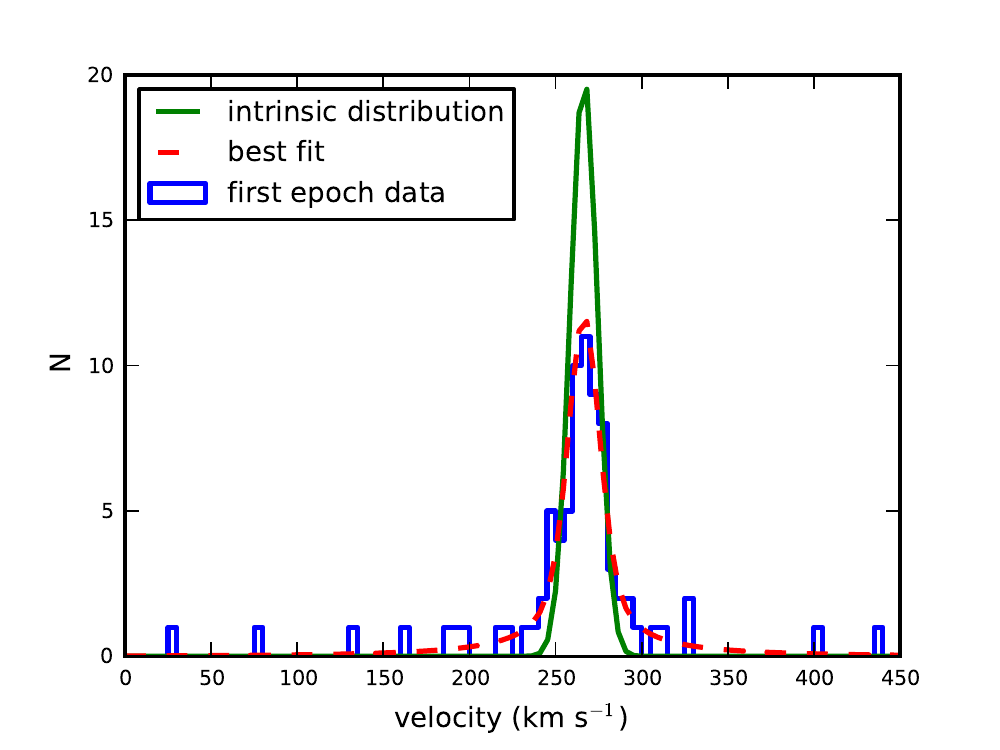}
    \caption{{\it Blue}: The observed radial velocity distribution for a single epoch drawn from the multi-epoch radial velocity dataset of R136 \citep[][see section \ref{data:R136} for details]{VHB2012, Sana_2013}. {\it Red dashed}: the best-fit radial velocity distribution, including the effect of the orbital motions of spectroscopic binaries. {\it Green}: The intrinsic velocity distribution of the cluster, corrected for the orbital motions of binaries. This intrinsic distribution is assumed to be a Gaussian whose width and mean have been optimized to fit the data (see section \ref{sec:method}). \label{fig:vel_fit}}
  \end{center}
\end{figure}

Finally the total likelihood $\mathcal{L}_i$ of observing a given velocity is obtained by summing the contributions from single stars (eq. \ref{eq:pdf_single}) and from binaries (eq. \ref{eq:pdf_bin}):
\begin{equation}
 \mathcal{L}_i = f_{\rm bin} \ {\rm pdf}_{\rm binary}(v_i, \sigma_i) + (1 - f_{\rm bin}) \ {\rm pdf}_{\rm single}(v_i, \sigma_i), \label{eq:L_1pop}
\end{equation}
where $f_{\rm bin}$ is the binary fraction and $1 - f_{\rm bin}$ is the fraction of systems that are single stars.

For a given set of binary orbital parameter distributions (section~\ref{sec:binaries}), we vary the binary fraction ($f_{\rm bin}$), the intrinsic line-of-sight velocity dispersion ($\sigma_v$) and the mean velocity ($\mu_v$) in order to find the maximum likelihood to reproduce all observed radial velocities (i.e. max$(\prod_i \mathcal{L}_i)$). We explore the probability distribution of the best-fit parameters and determine the uncertainties on each of these marginalized over all the other free parameters using a Metropolis-Hastings implementation of Markov Chain Monte Carlo (MCMC) simulations \citep[see][]{cottaar2012b}. 

Figure~\ref{fig:vel_fit} shows an example where we have fitted a single epoch of the multi-epoch radial velocities observed for R136 (for more details on this dataset see section~\ref{data:R136}). The binary orbital motions of the OB stars greatly affect the observed radial velocity distribution, as indicated by the large difference between the best-fit radial velocity distribution in red (given by equation~\ref{eq:L_1pop}) and the corresponding intrinsic Gaussian velocity distribution of the cluster in green (given by equation~\ref{eq:pdf_single}).

\subsection{Extension to multiple epochs\label{sec:extension}}
For a single epoch of radial velocity data, the maximum likelihood procedure described above gives the intrinsic velocity dispersion of a cluster corrected for the effect of binary orbital motions. This works well for stellar populations for which the binary orbital parameter distributions are well constrained \citep[e.g. solar-type stars,][]{cottaar2012b}. When these are more loosely constrained (e.g. for OB stars), multiple epochs can be used to identify and remove from the sample some spectroscopic binaries, thereby improving the accuracy of the final velocity dispersion determination. However, there will always remain undetected spectroscopic binaries, significantly affecting the radial velocity distribution. We describe below four adjustments to the likelihood procedure which make it possible to accurately estimate the intrinsic velocity dispersion of a cluster from a multi-epoch dataset corrected for the effect of these undetected spectroscopic binaries.

First, we have to identify as many spectroscopic binaries as possible and remove them from the sample. To identify these, binaries we use a $\chi^2$-test to assess if the radial velocity observations are consistent with coming from a single star displaying no radial velocity variations. For each star, we compute
\begin{equation}
\chi^2 = \sum_i \frac{(v_i - \bar{v})^2}{\sigma_i^2},
\label{chi2}
\end{equation}
where $\bar{v}$ is the weighted mean velocity over all the observed epochs $i$, and $v_i$ and $\sigma_i$ are the velocities at individual epochs and their uncertainties. By comparing this with the $\chi^2$-distribution (with the number of degrees of freedom equal to the number of epochs minus one), we compute the probability that a value as large as the computed $\chi^2$ can be due to measurement uncertainties alone. If this probability is low, we identify the star as a radial velocity variable and remove it from the sample. In line with previous work on R136 \citep{VHB2012}, we reject any star which has a probability $p < 10^{-4}$ of not being a radial velocity variable.

In principle, completely removing these radial velocity variables from the sample throws away information. However, including this information would require solving the radial velocity orbit of the binary. This is computationally very expensive if only a few epochs are available. Even if it is computed, the uncertainty in the systemic velocity of the binary will tend to be very weakly constrained, so it will provide little useful information about the intrinsic velocity dispersion or even the mean velocity of the cluster. Of course the systemic velocity of the binary should be included in the fit of the velocity dispersion (with its appropriate uncertainty), if enough epochs have been observed to solve the binary orbit.

Secondly, we compute, for the remaining and apparently single stars, what the effect of any unidentified binary might be on the observed radial velocity distribution. For this, we generate a large population of binaries sampled from the adopted binary orbital parameter distributions and assume that these systems have random phases and orientations. We then determine, using the same $\chi^2$-test as for the actual data, which binaries would have been detected given the timing of the observations and the measurement uncertainties. Eliminating these systems leaves us with a sample of binaries which would not be detected given the observational constraints. We compute the radial velocity offset distribution due to the orbital motions of these binaries, which gives us ${\rm pdf}_{\rm offset}$ for binaries which are undetected in the multi-epoch dataset. This has to be recomputed for every star, as the detectability of binaries depends strongly on the time sampling of the observations and the measurement uncertainties.

Our proposed method differs substantially from the multi-epoch fit from \citet{Martinez11}. These authors use a very similar likelihood procedure to fit their multi-epoch radial velocity data of the dwarf spheroidal galaxy Segue 1, however they do not make the explicit distinction between radial velocity variables and seemingly single stars. This distinction allows us to reach a larger precision, because the systemic velocities of the seemingly single stars have much smaller uncertainties than the systemic velocities of the radial velocity variables. By not distinguishing these populations, \citet{Martinez11} combine the probability distribution of the systemic velocity for seemingly single stars and radial velocity variables, which leads them to underestimate the constraints that the seemingly single stars can put on the intrinsic radial velocity distribution.

Having computed the radial velocity offsets due to binary orbital motions for the apparently single stars, we convolve this with an assumed intrinsic velocity distribution (equation \ref{eq:pdf_bin}) and add the contribution from truly single stars (equation \ref{eq:L_1pop}). However, when using equation \ref{eq:L_1pop} we have to keep in mind that the binary fraction among the apparently single stars ($f'_{\rm bin}$) will be lower than among the population as a whole:
\begin{equation}
 f'_{\rm bin} = f_{\rm bin} \frac{1 - f_{\rm det}}{1 - f_{\rm bin} f_{\rm det}}, \label{eq:new_fbin}
\end{equation}
where $f_{\rm det}$ is the fraction of binaries that would have been detected given the timing of the observations and the measurement uncertainties, as estimated from the random sample of binaries generated in the previous step.

Finally, we add an extra constraint on the binary fraction to match the observed fraction of binaries. This is given by a binomial distribution:
\begin{equation}
\mathcal{L} = \prod_{\rm variable} (f_{\rm bin} f_{\rm det}) \prod_{\rm not\ variable} (1 - f_{\rm bin}  f_{\rm det}), \label{eq:L_add}
\end{equation}
where we multiply the probability to detect a radial velocity variability for all stars showing radial velocity variability ($\prod_{\rm variable} (f_{\rm bin} f_{\rm det}) $) and the probability to detect no radial velocity variability for all apparently single stars ($\prod_{\rm not\ variable} (1 - f_{\rm bin}  f_{\rm det})$).

Here we briefly summarize this approach to fitting multi-epoch radial velocity data. Before performing the fit, four preparatory steps should be taken:
\begin{enumerate}
\item Separate the observed stars into radial velocity variables and seemingly single stars based on whether the radial velocity over the multiple epochs is consistent with being constant (see equation \ref{chi2}).
\item Draw a random population ($N \sim 10^5 - 10^6$) of binaries from the assumed period, mass ratio, and eccentricitry distribution and with a random phase and orientation.
\item For every observed seemingly single star, select a subsample of the simulated binary population by removing any simulated binary which would have been detected as a radial velocity variable given the time sampling and measurement uncertainty of that observed star. The fraction of radial velocity variables in the simulated population sets $f_{\rm det}$, which can be used to compute the probability that the observed seemingly single star is still a binary ($f'_{\rm bin}$) with equation \ref{eq:new_fbin}.
\item For every observed seemingly single star, compute the distribution of orbital velocities projected onto the line-of-sight for the pruned simulated binary population from step 3 in order to get ${\rm pdf}_{\rm offset}(v_i - v')$.
\end{enumerate}

We can then use the distribution of radial velocity offsets due to binary orbital motions to compute the likelihood of the data being drawn from a given intrinsic velocity distribution (e.g. parametrized by the mean velocity and velocity dispersion) and binary fraction by following the steps below. These steps will have to be repeated for every evaluation of the likelihood as part of the maximization procedure and associated analysis of the parameter uncertainties (e.g. from a Markov Chain Monte Carlo simulation).

\begin{enumerate}
\setcounter{enumi}{4}
\item For every observed seemingly single star, use the distribution from step 4, the probability of the star being a binary $f'_{\rm bin}$ from step 3, and an assumed underlying velocity distribution to compute the likelihood of reproducing the observed radial velocity of that star using equations \ref{eq:pdf_bin} and \ref{eq:L_1pop}.
\item Compute the total likelihood to reproduce the observed velocity distribution of seemingly single stars by multiplying the likelihoods to observe the radial velocities for all the observed seemingly single stars (or adding the log-likelihood).
\item Finally, multiply this likelihood from step 6 with the likelihood of detecting the observed fraction of radial velocity variables from equation \ref{eq:L_add} using the $f_{\rm det}$ from step 3.
\end{enumerate}

If only a single epoch of radial velocity data is available, steps 3-4 simplify to computing equation \ref{eq:pdf_offset} for the binary population from step 2. As in this case no radial velocity variables can be detected ($f_{\rm det} = 0$), the probability of a star to be a binary becomes the binary fraction (i.e. $f'_{\rm bin} = f_{\rm bin}$) and the likelihood from equation \ref{eq:L_add} will be equal to one for any binary fraction.

A possible extension of this procedure would be to perform a Bayesian analysis and allow the parametrization of the period, mass ratio, and eccentricity distributions to vary within the uncertainties set by previous observations. Although this would in principle accurately deal with the effect on the fitted velocity dispersion of the uncertainties in the binary orbital parameter distributions, it would require recomputing the preparatory steps 1-4 for every step in the optimization and subsequent Markov Chain Monte Carlo simulation. Unfortunately this is currently prohibitively slow, especially when multi-epoch data are considered.

\subsection{Constraints on the binary properties of OB stars\label{sec:binaries}}

\begin{table*}[!t]
\centering                 
\caption{Constraints on the distributions of orbital parameters of massive binaries from the literature. The intrinsic binary fractions inferred for these distributions of orbital parameters and their corresponding domain are listed. The intrinsic binary fractions corresponding to an extrapolation of the period distribution to 300 years are also given.}
\label{params}      
\begin{tabular}{l c c c c c c c}       
\hline\hline                 

Parameter & {\it pdf} & Domain & Variable & Value & Sample & Reference\\    
\hline
$\log_{10}$($P$/day)	& ($\log_{10} P$)$^{\pi}$	& 0.15 --- 3.5 & $\pi$ & -0.55$\pm$0.22& Galactic clusters &  \citet{Sana2012_science} \\
 & &0.15 --- 3.5    && -0.45$\pm$0.30 & 30 Doradus   &\citet{Sana_2013}  \\
  & & 0.0 --- 3.0  &  & 0.2$\pm$0.4 & Cyg OB2   &\citet{KK2012}  \\
\hline                       
$q = M_{2}/M_{1}$ & $q^{\kappa}$ & 0.1 --- 1.0& $\kappa$  & -0.1$\pm$0.6 & Galactic clusters &  \citet{Sana2012_science}\\    
  & & 0.1 --- 1.0 & & -1.0$\pm$0.4 & 30 Doradus&  \citet{Sana_2013}\\    
 & & 0.005 --- 1.0  &  & 0.1$\pm$0.5 & Cyg OB2  & \citet{KK2012} \\
  \hline                 
  e & $e^\eta$         & 0 --- 0.9 & $\eta$ &-0.45$\pm$0.17  & Galactic clusters &  \citet{Sana2012_science}\\  
    & & 0 --- 0.9 & & -0.5 (fixed) &   30 Doradus   & \citet{Sana_2013}  \\   
  & & 0.0001 --- 0.9 & & -0.6$\pm$0.3 &   Cyg OB2   &\citet{KK2012}  \\   
  \hline
  Binary fraction & & & & $69\pm9\%$ & Galactic clusters &  \citet{Sana2012_science}\\  
 & & & & $51\pm4\%$ &  30 Doradus   & \citet{Sana_2013}\\  
 & & & & $44\pm8\%$ & Cyg OB2   &\citet{KK2012}\\  
  \hline
  Binary fraction & & & & $85\pm11\%$ & Galactic clusters &  \citet{Sana2012_science}\\  
 for extrapolated& & & & $65\pm5\%$ &  30 Doradus   & \citet{Sana_2013}\\  
 period distribution& & & & $82\pm15\%$ & Cyg OB2   &\citet{KK2012}\\  
 \hline                                  
\end{tabular}
\end{table*}

\begin{figure*}[!t]
  \begin{center}
    \includegraphics[width=\textwidth]{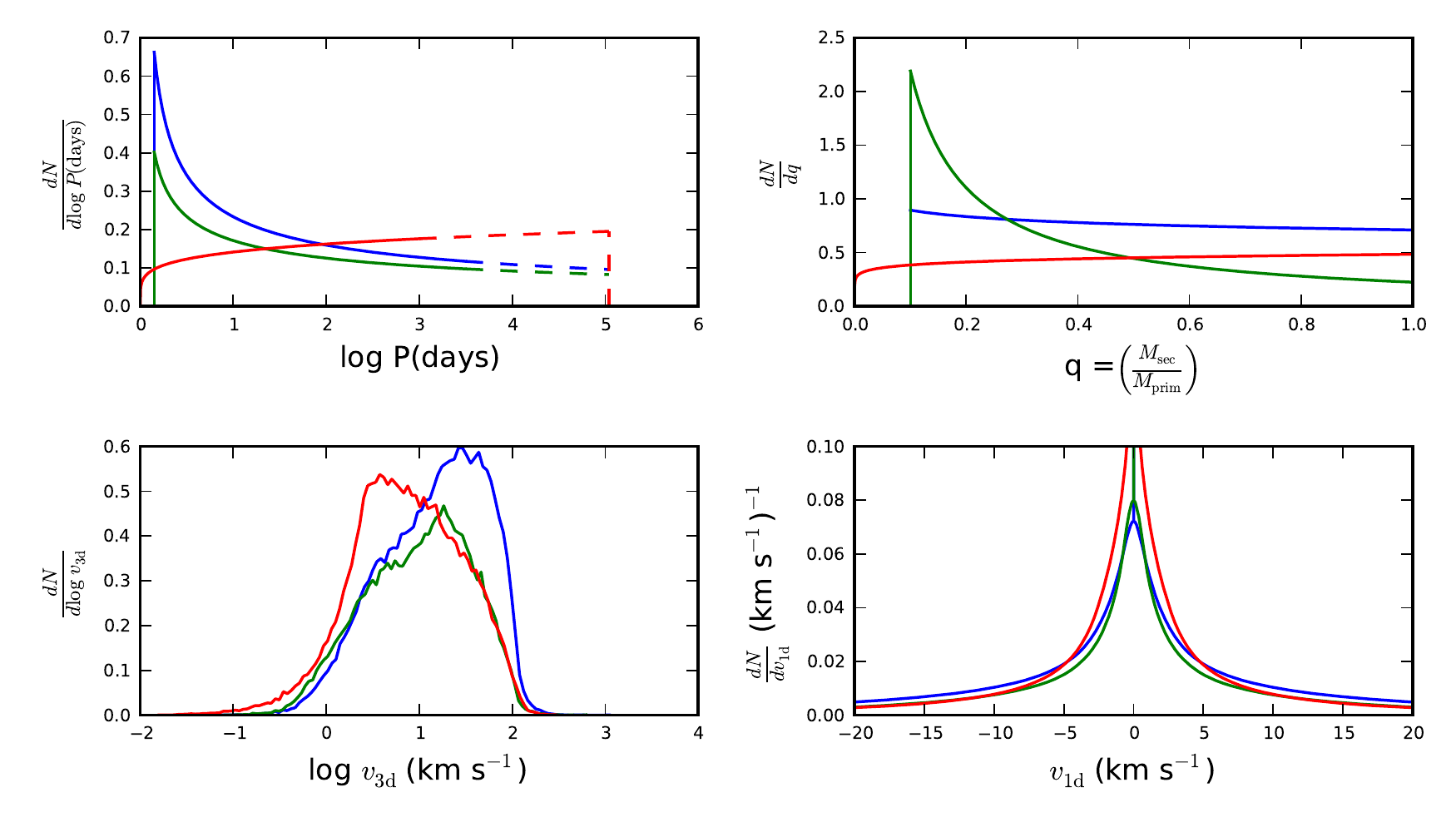}
    \caption{The assumed period distribution (upper left) and mass ratio distribution (upper right), as well as the unprojected logarithmic velocity distribution (lower left) and the projected velocity distribution (lower right). The extrapolation of the period distribution from the observed range to an upper limit of 300 years is plotted as a dashed line. This extrapolation has been taken into account in the lower panels showing the velocity distributions. The blue lines show the distributions from \citet{Sana2012_science}, the green lines from \citet{Sana_2013} and the red lines from \citet{KK2012}, as detailed in Table~\ref{params}. \label{fig:dist}}
  \end{center}
\end{figure*}

In what follows, we focus on the orbital parameter distributions of short-period close binaries, i.e. the ones which have the largest orbital velocities, the largest impact on the fitted line-of-sight velocity dispersion, and which are probed efficiently by Doppler shifts of spectral lines. As we review in Appendix~\ref{bin_review}, there is now ample evidence that massive stars are preferentially found in binary systems, in particular in these close spectroscopic binary systems.

Unfortunately, the orbital parameter distributions of these binaries are still relatively poorly constrained, despite their fundamental importance to star formation, stellar evolution, and the early dynamical evolution of massive star clusters. Even when studies of massive binaries include a relatively large number of systems \citep[e.g.][]{Garmany1980, Mason2009}, often only a small fraction of the identified binaries have well-constrained orbital properties \citep[see][]{SanaEvans2011}. A few recent studies, which we discuss below, have however achieved a better completeness in characterizing the identified binaries, in addition to correcting for observational biases in a more systematic way through Monte Carlo simulations (e.g. to translate the observed binary fraction into an intrinsic binary fraction).

\citet{KF2007} attempted to constrain the distributions of orbital parameters of massive binaries by using a sample of 900 radial velocity measurements of 32 O-type and 88 B-type stars in the Cyg OB2 association and comparing the raw velocities with the expectations of Monte Carlo simulations. They were however forced to make several simplifying assumptions about the orbital parameter distributions. Building upon this work and taking advantage of an extended dataset, \citet{KK2012} used 12 years of spectroscopic observations of 114 massive stars (B3--O5 primary stars) in Cyg OB2 and modelled the observed mass ratio, orbital period, and eccentricity distributions composed from the well-constrained orbital properties of 24 known binaries in the association (22 of which have periods shorter than 30 days; see caveats about this in Appendix~\ref{bin_review}). 

\citet{Sana_2013} analysed the multiplicity properties of the O-type star population of 30 Doradus through multi-epoch spectroscopy obtained as part of the VLT-FLAMES Tarantula Survey. With 360 O-type stars surveyed, this is the largest homogenous sample of massive stars analysed to date. However, given the limited number of epochs obtained (typically six), the orbital parameters could not be determined for individual binary systems. The intrinsic binary fraction and period and mass ratio distributions in this case were therefore constrained using Monte Carlo simulations to simultaneously reproduce the observed binary fraction, the distribution of the amplitudes of radial velocity variations and the distribution of the timescales of these variations. 

\citet{Sana2012_science} homogeneously analysed the O-type star population of six nearby Galactic open clusters and simultaneously fitted all the relevant intrinsic multiplicity properties. The larger average number of epochs allowed for a more complete binary detection compared to other studies. Over 75\% of the 40 binaries identified in this sample have measured orbital properties, which also made it possible for the authors to directly model and fit the orbital parameter distributions. More details about the studies mentioned in this section and other pioneering works are presented in Appendix~\ref{bin_review}.

\begin{figure}
  \begin{center}
    \includegraphics[width=.5\textwidth]{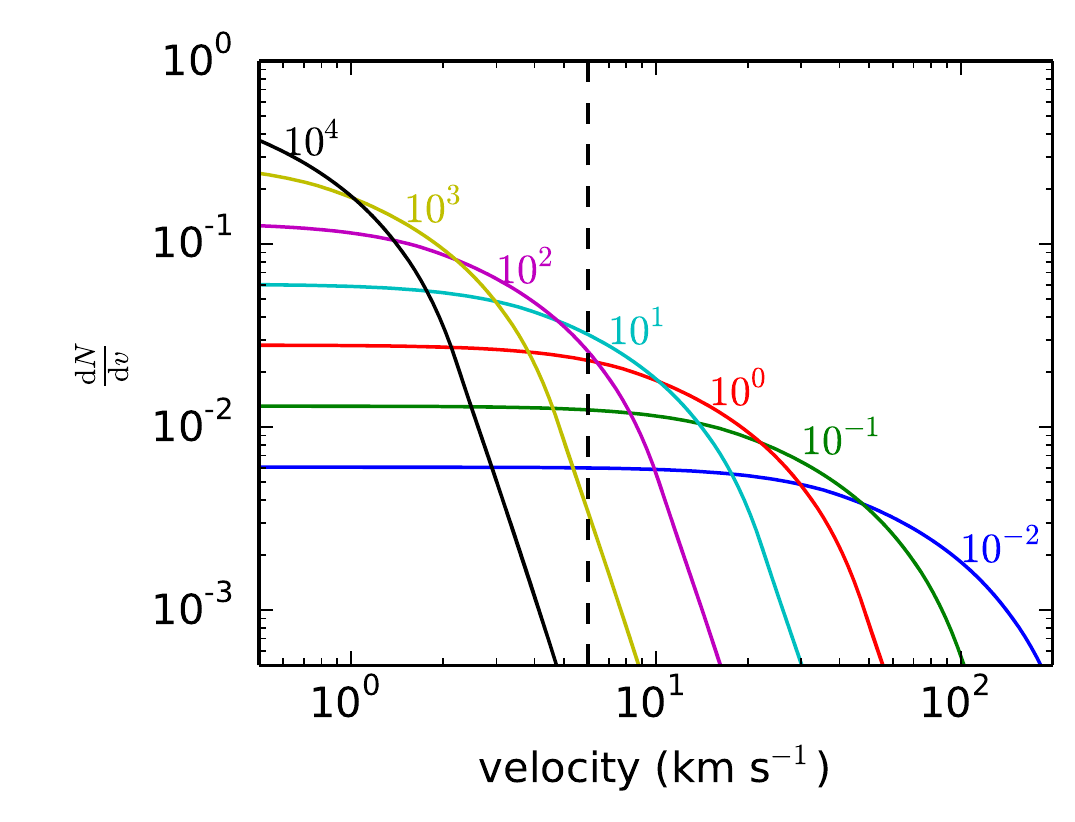}
    \caption{Probability distribution ($\frac{dN}{dv}$) for the radial velocity difference between the primary star in a binary and its centre of mass. Every line represents a binary with a fixed period and primary mass of 20 $\msun$, but random mass ratios, eccentricities, and phases drawn from flat distributions as well as a random orientation. The periods increase by factors of 10 and are labeled on the figure in units of years. The vertical dashed line marks the intrinsic velocity dispersion within 10~pc of R136 \citep[$\sim 6\ \kms$;][]{VHB2012}.\label{fig:period_velocity}  }
  \end{center}
\end{figure}

The distributions of binary orbital parameters from the three main studies discussed above are illustrated in Figure~\ref{fig:dist} and summarized in Table~\ref{params} along with the intrinsic spectroscopic binary fractions. The domains for which the period, mass ratio, and eccentricity have been considered and to which the quoted intrinsic binary fractions apply are also listed. These studies used power laws to describe the probability density functions of orbital periods (in $\log_{10}$ space), mass ratios and eccentricities with exponents $\pi$,  $\kappa$, and $\eta$, respectively. In
 the following sections, we will explore how sensitive the fitted velocity dispersion and binary fraction are to the adopted binary orbital parameter distribution by considering the differences between these orbital parameter distributions (which are consistent with each other at the 2-sigma level).

We focus on the binary properties from the three papers listed in Table~\ref{params} because these studies are based on large samples and corrected for observational biases in a systematic way through Monte Carlo simulations. In contrast to some other studies, they also have the advantage that they do not a priori assume a fixed distribution for the periods or mass ratios (apart from assuming a power-law functional form). Moreover, they sample different combinations of period and mass ratio distributions, cover a relatively wide range of values of $\pi$ and $\kappa$, and together are therefore representative of the current uncertainties on the binary properties of massive stars. These studies also each sample a different environment in terms of cluster mass and density. The clusters considered by \citet{Sana2012_science} have relatively low masses ($1\,000-5\,000\ \msun$), and it is currently unknown if the binary fraction and orbital parameter distributions should be affected in the more energetic environment around a $\sim10^5 \ \msun$ cluster like the 30 Doradus region, where the stars in the sample of \citet{Sana_2013} are located. Thus, while we may not expect all the scenarios presented in Table~\ref{params} to be applicable to the young massive cluster R136, it is interesting to test how the maximum likelihood method behaves under different assumptions about the binary properties.

In the studies discussed above, only binaries with a period of up to $\sim 10$ years could be detected due to the limited baselines of the observations. However, wider binaries still cause velocity offsets comparable to or larger than the velocity dispersion and hence significantly alter the observed velocity distribution. Figure \ref{fig:period_velocity} illustrates that large velocity offsets are likely to be caused by close binaries and small velocity offsets are more likely to be caused by wider binaries, as expected. Velocity offsets comparable to the velocity dispersion of R136 \citep[$\sim 6 \ \kms$; ][see also section~\ref{data:R136}]{VHB2012} are caused by binaries between $\sim 1$ and $\sim 100$ years, thus extending beyond the period range covered by spectroscopic surveys. To account for all the relevant binaries in our fits of the R136 data, we extrapolate the orbital period distribution to an upper limit of 300 years. We also increase the binary fraction to match the intrinsic binary fraction determined over the period range covered by the observations (last row in Table~\ref{params}).

\section{Monte Carlo simulations \label{sec:monte_carlo}}

\subsection{Data generation \label{sec:fake_data}}

We use Monte Carlo simulations to 
\begin{enumerate}
\item show the self-consistency of the method (i.e. ensure that it gives the correct result within the statistical uncertainties if all assumptions are met),
\item determine how the accuracy is limited due to the uncertainties on the orbital parameter distributions of massive binaries (section~\ref{sec:binaries}),
\item and determine how the precision is limited due to small-number statistics. 
\end{enumerate}
In these simulations, we create a mock dataset of radial velocities through a two-step procedure. First, we assign every star a systemic radial velocity from a Gaussian distribution with given mean velocity and intrinsic velocity dispersion. Then we assign an additional velocity representing the effect of the binary orbital motions to a subset of these stars (where the probability to be a binary is set by the binary fraction). These additional velocities are computed using binary orbital parameters randomly drawn from one of the period, mass ratio, and eccentricity distributions listed in Table \ref{params}. In all cases we use the extrapolated period distribution out to 300 years and the corresponding binary fraction (last row in Table \ref{params}). In these simulations, we do not look at the effect that measurement uncertainties or a non-Gaussian velocity distribution have on the best-fit parameters. The provided python code can be used to generate these radial velocity distributions and reproduce the Monte Carlo experiments performed in this paper.

These mock radial velocity datasets are fitted using our maximum likelihood procedure, assuming one of the three sets of binary orbital parameter distributions listed in Table~\ref{params}. This can either be the same set of orbital parameter distributions used to generate the data, allowing to test for self-consistency and constrain the precision of the method (goals 1 and 3 above), or a different set of orbital parameter distributions, allowing to test the accuracy of the procedure when the orbital parameter distributions in the cluster do not match those assumed (goal 2 above). We repeat these experiments for various sample sizes and intrinsic velocity distributions in the cluster.

In addition to these single-epoch Monte Carlo simulations, we ran a number of simulations where two epochs of radial velocity data have been sampled for every star. The radial velocity of the second epoch is generated by increasing the phase of the binary orbit with the ratio of the observational baseline $b$ and the binary period $P$ and recomputing the radial velocity offset from the centre of mass. The systematic centre-of-mass velocity of the binary is kept constant. 

Although in the single-epoch case we could safely ignore the measurement uncertainty (as long as it is sufficiently lower than the velocity dispersion such that it has a negligible effect on the observed velocity distribution\footnote{Even the case where the measurement uncertainty is too large for this approximation to be valid can be viewed as the measurement of a larger velocity dispersion $\sigma_{\rm v}^{\prime 2} = \sigma_{\rm v}^2 + \sigma_{\rm meas}^2$ with negligible measurement uncertainties, from which the (hopefully well-defined) measurement uncertainties are later subtracted.}), in the multi-epoch case the measurement uncertainty is crucial as it determines whether a specific binary can be spectroscopically detected. Thus, in this case, we add an additional (Gaussian) measurement uncertainty to every observation. For simplicity we adopt the same measurement uncertainty for all stars and epochs.

\subsection{Results}

\begin{figure}
  \begin{center}
    \includegraphics[width=.48\textwidth]{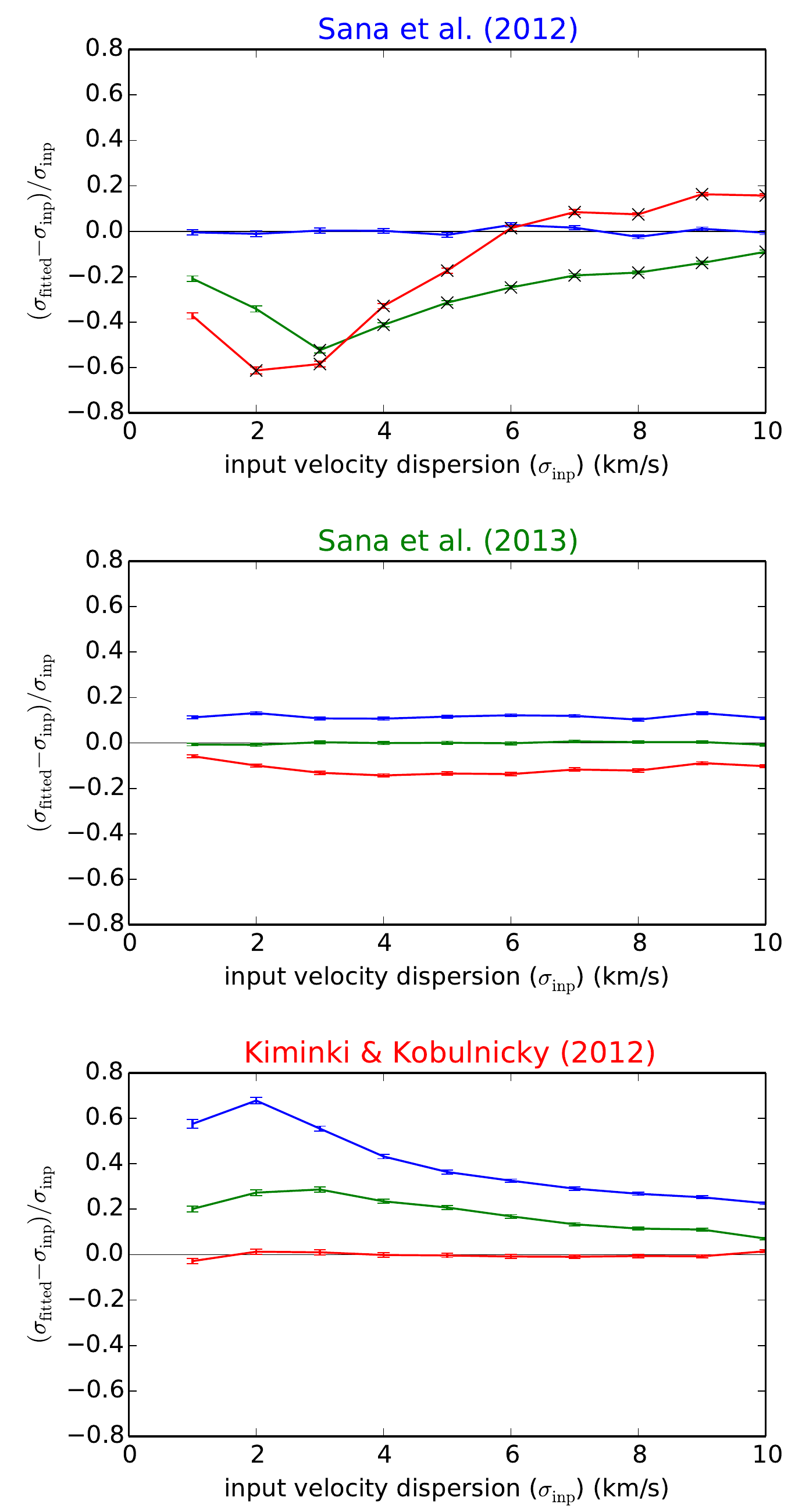}
    \caption{The systematic offset between the best-fit velocity dispersion ($\sigma_{\rm fitted}$) and the input velocity dispersion ($\sigma_{\rm inp}$) relative to the input velocity dispersion due to the uncertainties in the orbital parameter distributions of OB stars for a single epoch of radial velocity data. From top to bottom we have generated the radial velocity data using the orbital parameter distributions from \citet{Sana2012_science}, \citet{Sana_2013}, and \citet{KK2012}. The colors of the lines indicate the set of orbital parameter distributions assumed for the fit: blue for \citet{Sana2012_science}; green for \citet{Sana_2013}; red for  \citet{KK2012}. When the orbital parameter distributions used to generate the data are the same as those assumed in the fit, there is no systematic offset in the fitted velocity dispersion. The error bars represent the remaining uncertainty on the systematic offset after $\sim30$ Monte Carlo simulations, each of which included a sample of 2\,000 radial velocities. The simulations for which we find a best-fit binary fraction of 100\% in the majority of cases are crossed out in black (see upper panel) to indicate that the fitted velocity dispersion is likely to be overestimated in this case (see the main text). \label{fig:systematic}}
  \end{center}
\end{figure}

The uncertainties in the orbital parameter distributions of OB stars, encapsulated by the differences in the sets of distributions that we are considering, induces systematic offsets in the fitted velocity dispersion (and binary fraction) when an assumption about the underlying binary properties is made to fit the observed velocity distribution. We will first consider this systematic ofset in the velocity dispersion for single-epoch data and then consider how it changes if multi-epoch data are used. Figure~\ref{fig:systematic} shows the systematic offset in the fitted velocity dispersion induced when we fit the randomly generated radial velocity datasets (section~\ref{sec:fake_data}) assuming each of the sets of binary orbital parameter distributions listed in Table~\ref{params}.

In Figure \ref{fig:systematic}, we have crossed out the simulations for which we find a best-fit binary fraction of 100\%. Such a high best-fit binary fraction indicates that there are more high velocity outliers in the dataset than can be explained by the assumed binary orbital parameter distributions. These high-velocity outliers could have many causes in real datasets, such as ejected stars, contaminating field stars, or bad radial velocity measurements. In the case of the top panel of Figure \ref{fig:systematic}, the radial velocity datasets were generated using the binary properties from \citet{Sana2012_science}, which includes a relatively large fraction of close and similar-mass binaries. This leads to many high-velocity outliers in these datasets, which cannot be matched by the other orbital parameter distributions considered here (even for a binary fraction of 100\%). This situation, where the assumed orbital parameter distributions cannot explain the large number of high-velocity outliers, will cause the fitted velocity dispersion to be systematically overestimated, as this reduces the number of outliers. So irrespective of the nature of the high-velocity outliers, when a binary fraction of 100\% is found the fitted velocity dispersion will tend to be inflated and be unreliable. Therefore, below we will only consider the cases where we find a best-fit binary fraction below 100\%.

When we use the same orbital parameter distributions to fit the data as we used to generate the data, no systematic offset between the input and retrieved velocity dispersion are found (Figure \ref{fig:systematic}; blue in the upper panel, green in the middle panel, red in the bottom panel). From this we conclude that the procedure is self-consistent (i.e. it yields the correct result if the assumptions made for the fit match the properties of the generated dataset). Significant systematic offsets are found when different orbital parameter distributions are used to generate the data than to fit the data. These systematic biases in the fitted velocity dispersion range from up to 60\% for small input velocity dispersions ($\sim 2\ \kms$) to as low as 25\% for velocity dispersions of $10\ \kms$ (other lines in Figure \ref{fig:systematic}) for the sets of orbital parameter distributions under study here. This suggests that without precise knowledge of the binary properties of massive stars in a given environment, we are limited to an accuracy of tens of percent when determining the line-of-sight velocity dispersion using a single epoch of radial velocities from OB stars. Most OB stars will form in massive clusters, whose virial velocity dispersion is generally expected to be above $4\ \kms$. In this range the velocity dispersion can be fitted to an accuracy better than $\sim 40\%$, which allows a cluster in virial equilibrium to be barely distinguished from an unbound cluster from a single epoch of data. 

\begin{figure}[!t]
  \begin{center}
    \includegraphics[width=.49\textwidth]{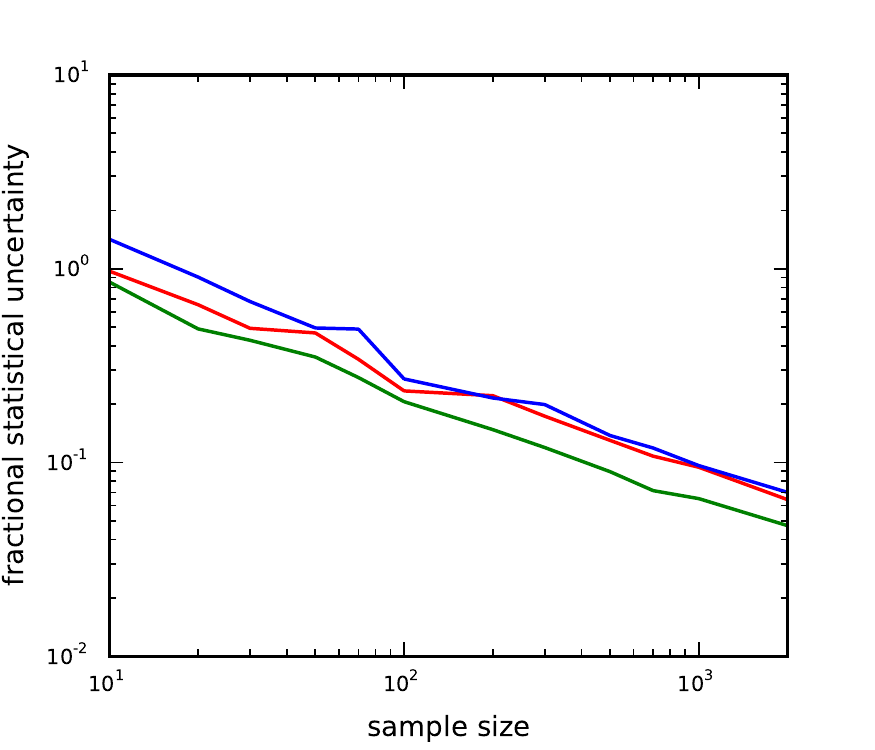}
    \caption{The 1-sigma statistical uncertainties due to small-number statistics in the fitted velocity dispersion as a function of the sample size $N$ for single-epoch observations of a cluster with a velocity dispersion of 5~$\kms$. The colors indicate the set of orbital parameter distributions used to generate the data and assumed for the fit: blue for \citet{Sana2012_science}; green for \citet{Sana_2013}; red for  \citet{KK2012}. The color coding is the same as in Figure \ref{fig:systematic}. The statistical uncertainties are given relative to the input velocity dispersion of 5~$\kms$. Every point in this graph represents the average statistical uncertainty (as measured by the MCMC) in 60 Monte Carlo simulations, except for sample sizes less than or equal to 30 where 200 Monte Carlo simulations were averaged.\label{fig:random}}
  \end{center}
\end{figure}

If it is larger than the statistical uncertainties due to small-number statistics, this systematic offset of $\sim40\%$ will limit the accuracy on the velocity dispersion. Figure \ref{fig:random} shows how the statistical uncertainties of the fitted velocity dispersion depends on the sample size. From this figure we find that a 2\,$\sigma$ statistical uncertainty of $\sim40\%$ is reached for a sample size of $\sim100$ stars (for a cluster with a velocity dispersion of $5 \ \kms$). For a larger sample size the systematic offset due to the loosely constrained binary orbital parameter distributions will dominate the uncertainty on the velocity dispersion.

\begin{figure}
  \begin{center}
    \includegraphics[width=.53\textwidth]{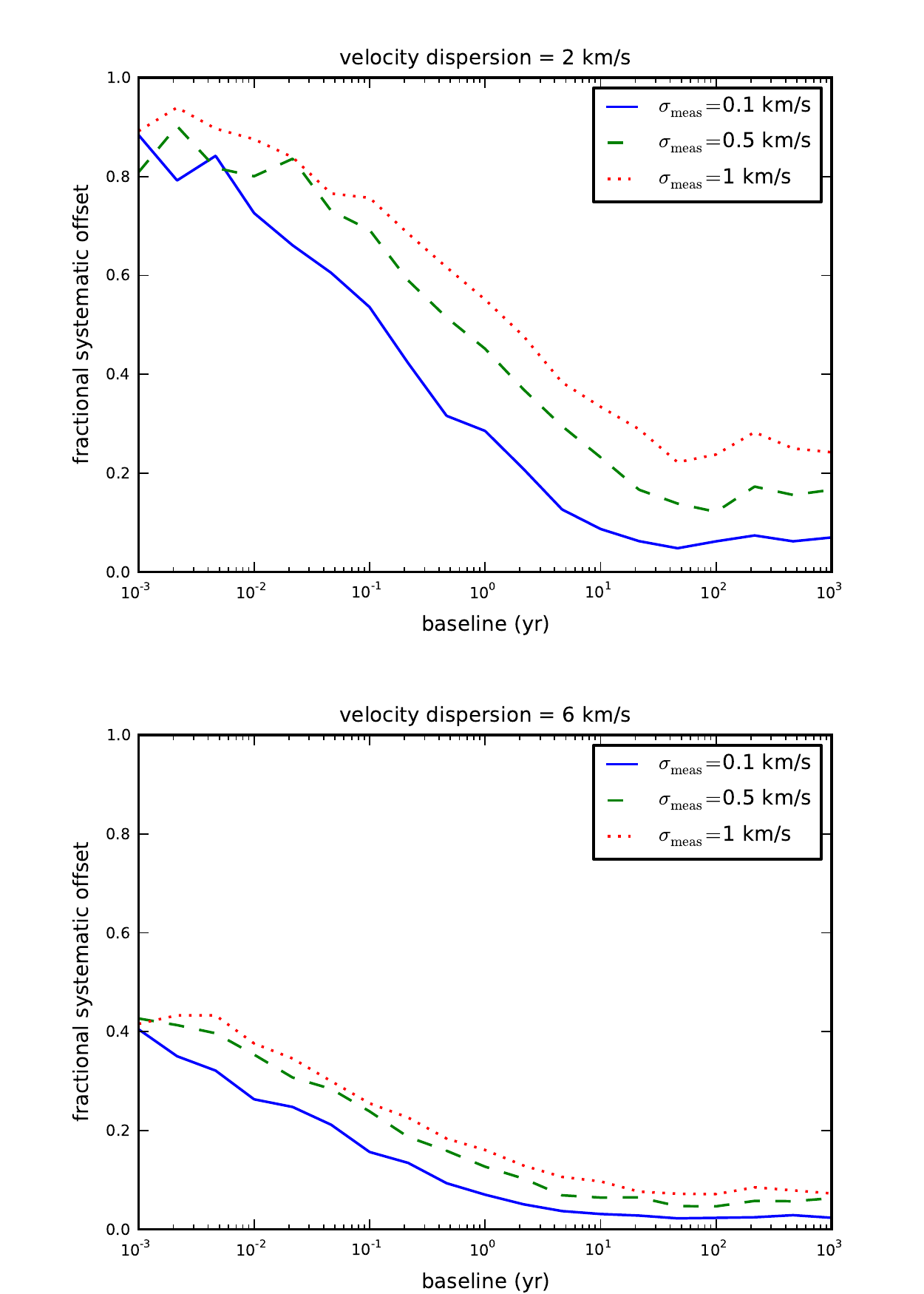}
    \caption{The systematic offset in the fitted velocity dispersion for radial velocity data generated using the \citet{KK2012} binary orbital parameter distribution and fitted assuming the \citet{Sana2012_science} binary orbital parameter distribution, plotted against the baseline between two epochs of observations. As the baseline between the two epochs of observations increases, more spectroscopic binaries are detected, leading to a decrease in the systematic offset of the velocity dispersion. We show the results from Monte Carlo simulations for clusters with a velocity dispersion of 2 $\kms$ (upper panel) and 6 $\kms$ (lower panel), as well as for measurement uncertainties of 0.1 $\kms$ (solid), 0.5 $\kms$ (dashed), and 1 $\kms$ (dotted).\label{fig:2epoch}}
    \end{center}
\end{figure}

\begin{figure}
  \begin{center}
    \includegraphics[width=.53\textwidth]{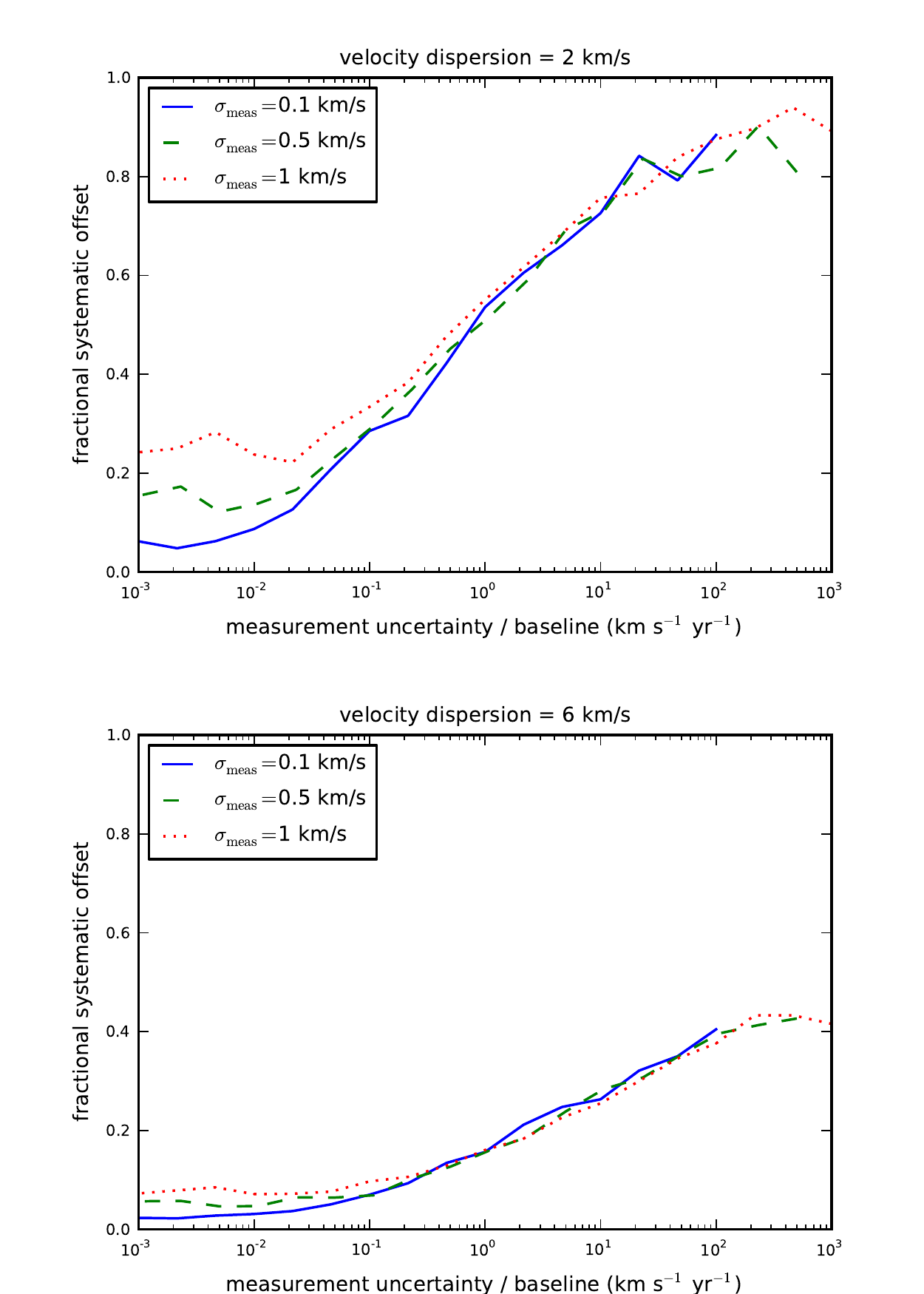}
    \caption{Same as Figure \ref{fig:2epoch}, but with measurement uncertainty over the baseline ($\sigma_{\rm meas}/b$), rather than the baseline on the x-axis. Note that for a broad region of intermediate baselines, all the curves overlap. \label{fig:2epoch_norm}}
    \end{center}
\end{figure}

One option to decrease the systematic offset in the fitted velocity dispersion is to detect the spectroscopic binaries which alter the observed velocity distribution. This requires multiple epochs of data. The largest systematic offset was found when radial velocity data generated using the binary orbital parameter distributions from \citet{KK2012} was fitted using the distributions from \citet{Sana2012_science} (blue line in lower panel of Figure~\ref{fig:systematic}). For this specific case, we explored whether the systematic offset in the fitted velocity dispersion could be resolved by using two epochs of radial velocity data. The results of these Monte Carlo simulations are shown in Figures \ref{fig:2epoch} and \ref{fig:2epoch_norm} for clusters with a velocity dispersion of $2 \ \kms$ and $6 \ \kms$, for measurement uncertainties $\sigma_{\rm meas}$ of 0.1, 0.5, and 1~$\kms$ and a broad range of baselines between a few hours and hundreds of years. 

A larger fraction of the spectroscopic binaries can be detected when the baseline is longer or the measurement uncertainties are smaller. As expected, this will lead to a smaller systematic offset in the fitted velocity dispersion. From Figure \ref{fig:2epoch_norm} we find that for a broad range of intermediate baselines the systematic offset decreases linearly with $\log(\sigma_{\rm meas}/b)$, such that the systematic offset decreases by the same amount for every doubling of the baseline or halving of the measurement uncertainties. The size of the offset and the rate of decline depend on the exact binary properties assumed to compute the offset. This relation break down both for low $\sigma_{\rm meas}/b$, when only very close binaries can be identified (whose large velocity offsets already marked them as binaries in single-epoch data), as well as for high baselines $b$, when the baseline becomes comparable to the period of the widest binaries, that have a significant radial velocity offset. 

In section~\ref{sec:multiepoch}, we analyse these relations in more detail. Here we note that for the typical measurement uncertainties of 4 $\kms$ and the total baseline of 1 year of the radial velocity observations in R136, we get $\sigma_{\rm meas}/b \approx 4 \ \kms {\rm yr}^{-1}$. So from these Monte Carlo simulations we expect a significant systematic offset in the fitted velocity dispersion of $\sim 40\%$ depending on the adopted binary orbital parameter distribution, when analyzing a single epoch of the R136 radial velocity data. This systematic offset should be significantly reduced to $\sim 20\%$ when considering the full baseline of one year of the radial velocity observations (see lower panel of Figure \ref{fig:2epoch_norm} for $\sigma_{\rm meas}/b \approx 4 \ \kms {\rm yr}^{-1}$).

\section{R136: a young massive cluster \label{R136}}

\subsection{Data \label{data:R136}}

The dataset that we use to test the maximum likelihood method outlined in section~\ref{sec:method} consists of multiple epochs (at least five) of radial velocity measurements for 81 O-type systems in the inner 10~pc (in projection) of R136, all obtained with the FLAMES instrument on the VLT as part of the VLT-FLAMES Tarantula Survey (VFTS). These 81 systems include both apparently single stars and objects showing radial velocity variability. Most of the stars in the inner 5~pc have been observed with the ARGUS integral-field unit coupled to the Giraffe spectrograph, while for the vast majority of stars between 5 and 10~pc the Medusa fibre-feed to Giraffe has been used \citep[for more details on the VFTS data, see][]{VFTS_I}. In both cases, the radial velocities were measured by fitting Gaussians to helium absorption lines using a similar approach and the same rest wavelengths \citep{VHB2012, Sana_2013}. This sample of 81 objects excludes the B-type and emission-line stars observed by the VFTS in the inner 10~pc of R136, but it includes a few supergiants (or supergiant candidates) for which the absolute radial velocities might be inaccurate due to the effect of stellar winds on the line profiles. These were however shown to have a negligible impact on the measured velocity dispersion of the apparently single stars of the sample \citep[see][]{VHB2012}. The radial velocity measurements for individual epochs are listed in \citet{VHB2012} and \citet{Sana_2013}. The median radial velocity uncertainty of the single-epoch measurements for the objects of our sample is $\sim4 \ \kms$. Note that we do not include stars observed by the VFTS further out than 10~pc to limit possible contamination from nearby clusters or other star formation events in the surroundings of R136. 

\begin{figure}[!t]
  \begin{center}
    \includegraphics[width=.5\textwidth]{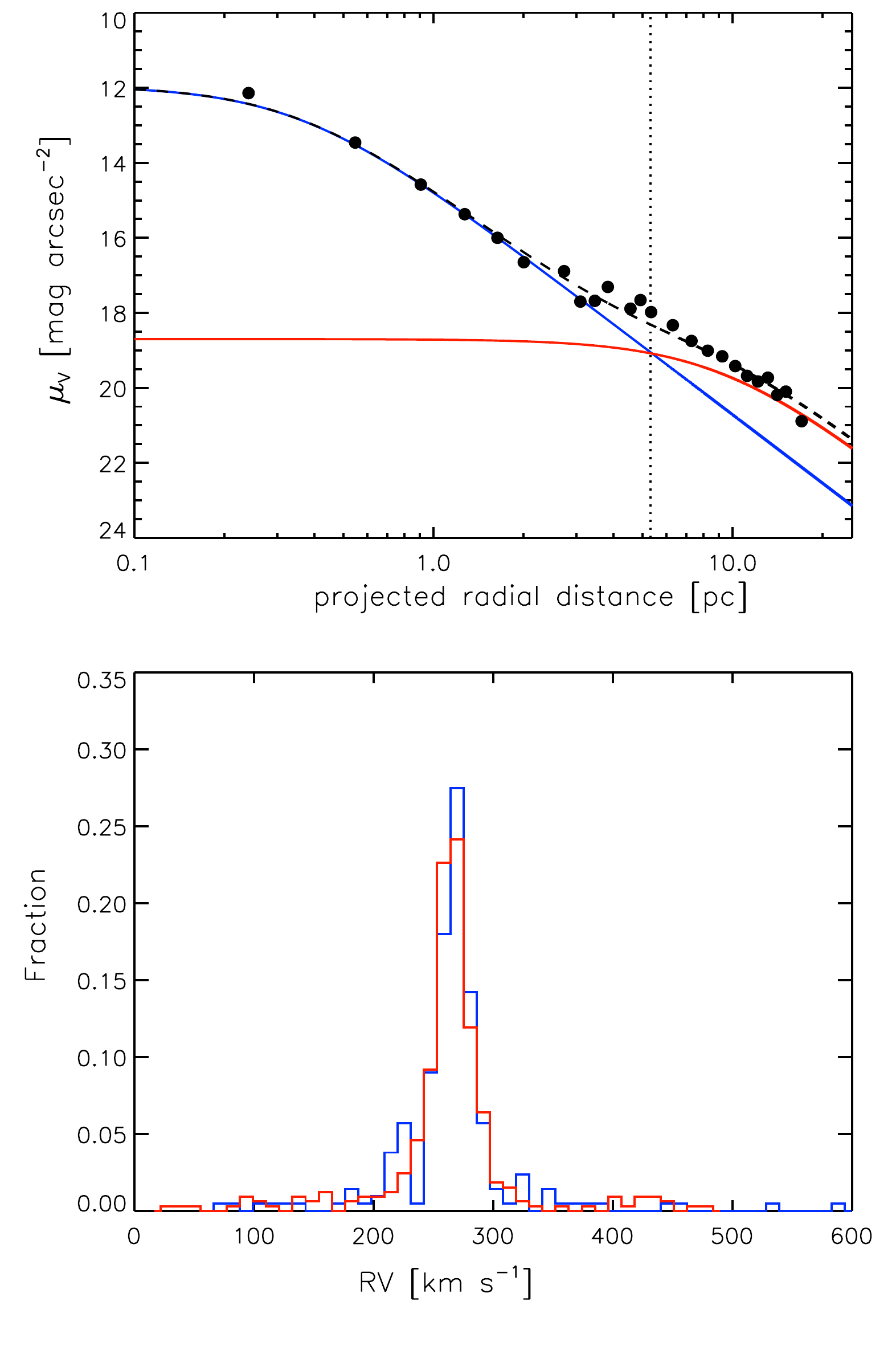}
    \caption{{\it Top:} Two-component EFF fit to the surface brightness profile of R136 as in \citet{mackey2003}. The data points shown are Mackey and Gilmore's recalibrated surface brightnesses from \citet{mclaughlin2005}. Stars that are further out than $\sim5$~pc (marked by the dotted line) from the cluster centre are more likely to be part of the OB association. Within this radius stars are more likely to belong to the R136 cluster itself. {\it Bottom:} Normalized histograms showing the velocity distribution of the 32 stars within 5~pc in blue (211 velocities), and the 49 stars between 5 and 10~pc in red (327 velocities). \label{fig:two_comp}}
  \end{center} 
  \end{figure}

\citet{mackey2003} found that the light profile of R136 is best fitted with a double-component EFF profile \citep{EFF}, suggesting that the cluster is superimposed on an OB association contributing to a significant fraction of its total integrated light \citep{maiz2001}. In this double-component EFF fit, the projected radius where the two components contribute equally is at about 5 pc, as illustrated in Figure~\ref{fig:two_comp} (top panel). To check if the velocity distributions of these two potentially distinct components should be considered separately, we also show in Figure~\ref{fig:two_comp} (bottom panel) the velocity distribution of the stars more likely to be part of the inner component (i.e. the cluster) compared to the velocity distribution of the stars more likely to be part of the outer component (i.e. the OB association). We see no evidence for a difference in the velocity distribution between the two components. Similarly, \citet{VHB2012} found the measured velocity dispersion profile to be relatively flat between 1 and 10~pc from the centre of R136\footnote{Even though only the inner $\sim5-6$~pc are more than a crossing time old and can be considered to be bound, and thus strictly speaking as part of the cluster according to the definition of \citet{gielesPZ2011}.}. Because the two components appear indistinguishable (at least kinematically) in the inner 10~pc, we conclude that it is justified to try to fit a single velocity dispersion for all the stars of our sample.

Note that \citet{sabbi2012} found a dual structure in the density of low mass stars in R136, hinting at a merger event between the main core of R136 and a second cluster. Because only a very small fraction our targets are expected to be associated with this second clump, we treat all the stars within 10~pc as one population.

\subsection{Results}

\citet{VHB2012} determined a velocity dispersion of $\sim6 \ \kms$ for the O-type stars within 10~pc in projection from the centre of R136\footnote{The velocity dispersion was slightly smaller ($\sim5 \ \kms$) when considering only the stars within 5~pc from the centre. We ignore the small potential contribution ($\sim0.5\ \kms$) to the velocity dispersion from cluster rotation \citep{vhb2012_rot}.} after selecting out identified binaries and estimating the effect of undetected binaries. To test our procedure, we fit single-epoch radial velocity datasets extracted from the full multi-epoch dataset presented in section~\ref{data:R136}. One such fit is shown in Figure \ref{fig:vel_fit}.

We extract five single-epoch radial velocity datasets without any overlap because every star in the sample has been observed for at least five epochs. These datasets are not independent, as they all include observations of the same stars, albeit at different epochs.

\begin{figure*}
  \begin{center}
    \includegraphics[width=1.\textwidth]{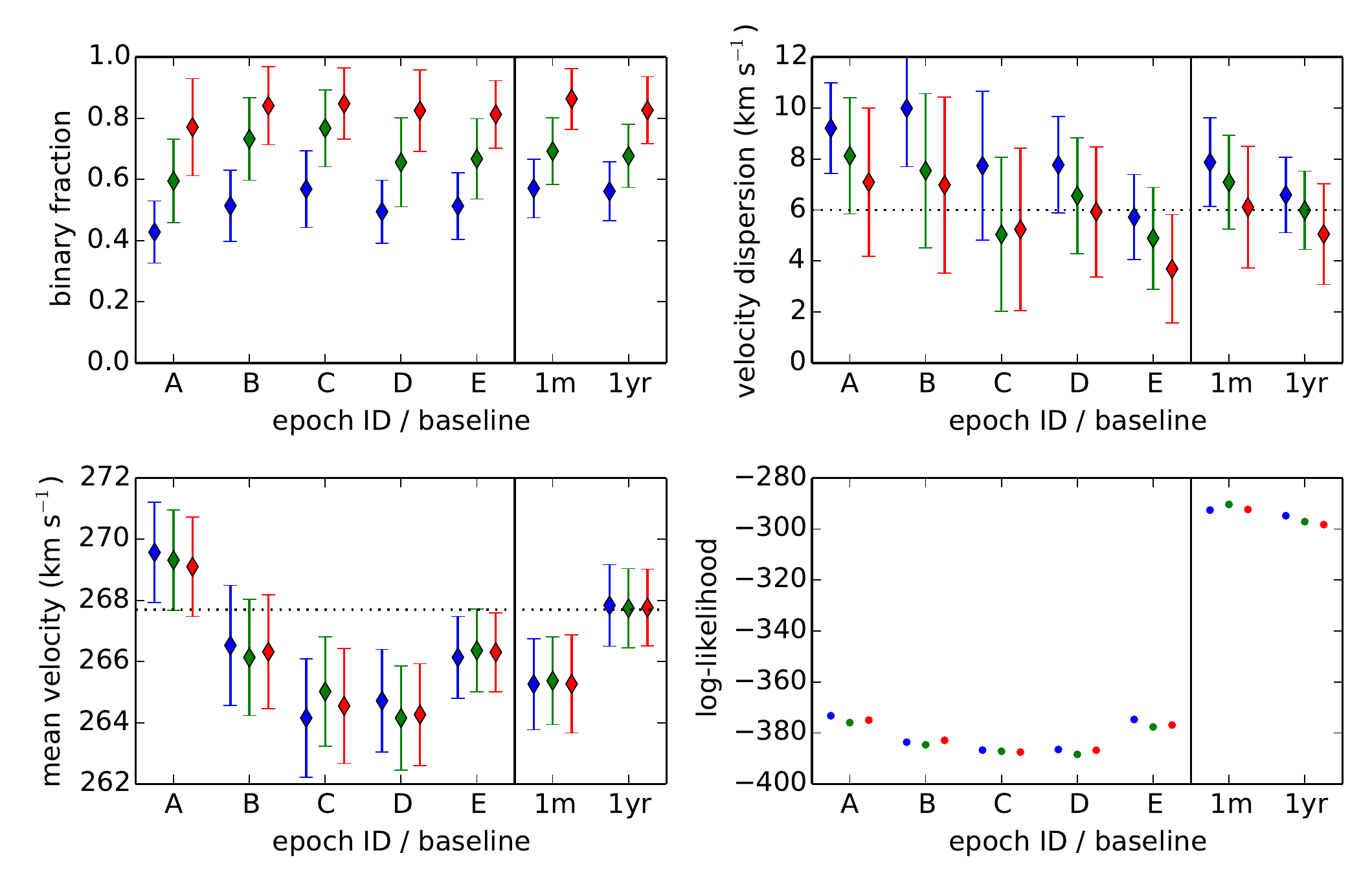}
    \caption{The best-fit binary fraction, velocity dispersion, and mean velocity as well as the log-likelihood of the best fit for five single-epoch radial velocity datasets (labeled A-E) of the O-type stars in R136, as well as two datasets with two epochs with a baseline of roughly one month and one year. The colors indicate the set of orbital parameter distributions assumed in the fit: blue from \citet{Sana2012_science}; green from \citet{Sana_2013} ; and red from \citet{KK2012}. The error bars show the 1\,$\sigma$ uncertainties on the best-fit values, as estimated through Markov Chain Monte Carlo simulations. The dashed lines in the upper-right and lower-left panels show the values found after removing the spectroscopic binaries identified from the full multi-epoch dataset \citep{VHB2012}. \label{fig:R136_fit}}
  \end{center}
\end{figure*}

The best-fit binary fraction, velocity dispersion and mean velocity for each of these five datasets are shown on the left side of each panel in Figure \ref{fig:R136_fit}. The three different colors again represent the three sets of orbital parameter distributions assumed. The 1\,$\sigma$ error bars are somewhat larger than predicted from the Monte Carlo simulations for a sample of 81 radial velocities (see Figure~\ref{fig:random}), because the effective sample size is smaller than 81. A significant subset of the observed radial velocities in R136 have a measurement uncertainty comparable to or larger than the velocity dispersion. These radial velocities provide (nearly) no information about the velocity dispersion of the clusters, leading to a smaller effective sample size.

We find lower binary fractions for the orbital parameter distributions where an individual binary has a larger likelihood of causing a significant radial velocity offset. This explains the low binary fraction found for the parameters from \citet[][blue in Figure \ref{fig:R136_fit}]{Sana2012_science}, which has relatively many close and similar-mass binaries (see Table \ref{params} and Figure~\ref{fig:dist}). In contrast, the best-fit mass-ratio distribution of \citet{Sana_2013} has more low-mass binary companions, while the period distribution favoured by \citet{KK2012} has a larger proportion of wide binaries and hence requires a larger binary fraction to reproduce the same number of high-velocity outliers.

The fitted velocity dispersion is less sensitive to the exact parameters assumed than the binary fraction, although we do still find variations of $\sim 3 \ \kms$ (see central panel in Figure~\ref{fig:R136_fit}) for the single epoch fits, which is comparable to the $\sim40\%$ systematic offset found from Monte Carlo simulations (section~\ref{sec:monte_carlo}). The intrinsic velocity dispersion of R136 fitted from single-epoch radial velocities is consistent with the $6\ \kms$ from the multi-epoch approach \citep{VHB2012}.

Finally the best-fit mean velocity is hardly sensitive to the assumed orbital parameter distributions (see Figure~\ref{fig:R136_fit}), because as long as the projections of the binaries on the sky are random, the velocities from the binary orbital motions will be symmetrically distributed around the mean velocity. So the systematic uncertainties induced from an assumed binary population does not significanlty affect measurements of the mean velocity, which means that parameters depending on the mean velocity (e.g. cluster rotation) can be accurately fitted.

The best-fit parameters obtained when using two epochs of data are shown on the right side of each panel of Figure~\ref{fig:R136_fit}. As already seen from the Monte Carlo simulations of section \ref{sec:monte_carlo}, the systematic offset in the fitted velocity dispersion induced when different binary orbital parameter distributions are used decreases significantly with multi-epoch data. For the single-epoch data, we found typical variations in the fitted velocity dispersion of $\sim 3 \ \kms$, which decreases to only $\sim 2\ \kms$ when using two epochs which are approximately one month apart. When a baseline of one year is used, the systematic offset decreases even further to about $1\ \kms$.

Although the fitted velocity dispersions show smaller systematic offsets between adopted binary orbital parameter distributions when multi-epoch data are used, the fitted binary fraction is roughly the same for single- and multi-epoch data. This is consistent with the fitted binary fraction being driven by the proportion of close and similar-mass binaries. When assuming a high proportion of such systems, the fitted binary fraction will be lower, irrespective of whether these stars appear as high-velocity outliers (as in the single-epoch case) or as radial velocity variables (as in the multi-epoch case).

From inspecting the log-likelihood of the best-fit models for R136 (Figure~\ref{fig:R136_fit}), no set of orbital parameter distributions \citep{Sana2012_science, Sana_2013, KK2012} consistently fits the data better than the others. This does not imply that there is insufficent information in the dataset to distinguish between the orbital parameter distributions. After all our method ignores some information (e.g. the observed distribution of radial velocity offsets between epochs) in the log-likelihood, which could have been used to constrain the binary orbital parameter distributions \citep[e.g.][]{Sana_2013}, but not the intrinsic velocity dispersion.

\section{Discussion \label{discussion}}
We are able to fit the intrinsic velocity dispersion from a single epoch of radial velocity data with an accuracy ranging from $\sim 60\%$ for a cluster velocity dispersion of $\sim 2\ \kms$ to $\sim 25\%$ for $\sigma_v \sim 10\ \kms$, given the present-day uncertainties on the orbital parameter distributions of OB spectroscopic binaries (e.g. Table~\ref{params} and Figure~\ref{fig:dist}). For typical velocity dispersions of young massive clusters ($\gtrsim 4 \ \kms$) we find that an accuracy better than 40\% can be reached, which is sufficient to distinguish a cluster in virial equilibrium from an unbound cluster. This systematic offset will dominate over the statistical uncertainties from small-number statistics for a sample size larger than about 100 stars. This means that after observing 100 stars for a single epoch, the accuracy of the fitted velocity dispersion will in principle no longer increase when more stars are observed due to the uncertainties in the binary orbital parameter distributions of OB stars. Observing the same stars for multiple epochs makes it possible to push down the systematic offset. For example, a ratio of the measurement uncertainty over the baseline of $1\ \kms {\rm yr}^{-1}$ is required to push the systematics below 15\% for an intrinsic velocity dispersion of 6~$\kms$,

To understand the systematic offsets, we directly compare the fits of the single-epoch R136 data for the orbital parameter distributions of \citet{Sana2012_science} and \citet{KK2012}. In the case of \citeauthor{Sana2012_science}, the binaries typically orbit each other with tens of $\kms$, an order of magnitude larger than for \citeauthor{KK2012} (Figure~\ref{fig:dist}). Thus, an individual binary in the former case will have a much large probability of causing a high-velocity outlier. To reproduce the high-velocity outliers in the R136 data we thus need a lower binary fraction for the binary orbital parameter distribution of \citet{Sana2012_science} than for \citet{KK2012}, which explains the difference in the binary fraction found for the R136 data (left panel in Figure \ref{fig:R136_fit}).

On the other hand, the fitted velocity dispersion is affected by the abundance of wider binaries with orbital velocities comparable to the intrinsic velocity dispersion of the cluster ($\sim 6 \ \kms$), as these binaries are more likely to really broaden the observed peak in the velocity distribution, instead of just creating a high-velocity tail. We can see in Figure \ref{fig:dist} that these binaries are much more common for the orbital parameter distributions of \citet{KK2012} than for those of \citet{Sana2012_science}, even before we take into account the fact that we find a lower overall binary fraction for \citet{Sana2012_science}. This much higher fraction of wide binaries broadening the observed velocity distribution when assuming the binary orbital parameter distributions of \citet{KK2012} leads to a significantly smaller fitted velocity dispersion (central panel in Figure \ref{fig:R136_fit}) compared to \citet{Sana2012_science}. Still, it is remarkable that the systematic offset in the velocity dispersion is smaller than a factor of 2 given that the peaks in the orbital velocity distributions differ by an order of magnitude (see lower left panel in Figure~\ref{fig:dist}).

Note that the differences between the sets of binary properties listed in Table~\ref{params} might be real to some extent (i.e. not only reflect uncertainties in our knowledge of these properties). There are reasons to believe that the binary properties of massive stars might depend on the environment and the age of the region in which they are located. For example, the intrinsic fraction of O-type spectroscopic binaries seems lower in 30 Doradus than in the relatively low-density Galactic clusters (see Table~\ref{params}), although both results still agree within 2\,$\sigma$. As discussed by \citet{Sana_2013}, this might suggest that the binary properties in the 30 Doradus region have already been significantly affected by dynamical and/or stellar evolution which would induce merger events or binary disruption and decrease the observed number of binaries. This is to be expected given the presence of different populations in the region, some already quite old, and the fact that a fraction of the O-star population consists of runaways. For the Galactic open clusters sample, cluster dynamics and stellar evolution are not expected to have significantly altered the orbital properties of the binaries \citep{Sana2012_science}. Given the young age of the clusters in this sample, the parameters reported in this case are probably a good representation of the properties of massive binaries at birth.

As discussed in Appendix~\ref{bin_review}, a wide variety of possible period, mass ratio, and eccentricity distributions for OB binaries have been considered in the literature. We have chosen to limit our analysis to the three sets of orbital parameter distributions in Table~\ref{params}, which we felt were the most representative of our present-day knowledge about these distributions. However, the reader is invited to use the provided python code\footnote{https://github.com/MichielCottaar/velbin}, which can be used to fit an observed radial velocity distribution, as well as create and fit mock radial velocity distributions, using either the orbital parameter distributions described in this work or any other orbital parameter distributions. This code can thus be used to calculate the systematic offsets in the fitted velocity dispersion between any orbital parameter distributions (as in Figure \ref{fig:systematic}) or to determine the accuracy with which the velocity dispersion can be fitted as a function of sample size when planning observations (as in Figure \ref{fig:random}).

\subsection{The effectiveness of multi-epoch radial velocities\label{sec:multiepoch}}
Multi-epoch data is required to reduce the systematic offsets in the fitted velocity dispersion due to uncertainties in the adopted binary orbital parameter distributions. In particular, one needs to identify and remove from the sample those spectroscopic binaries with radial velocity offsets comparable to the intrinsic velocity dispersion. Binaries causing a larger velocity offset are already easily identified in single-epoch data as being high-velocity outliers and hence have little effect on the fitted velocity dispersion. Binaries causing a velocity offset smaller than the velocity dispersion have a negligible impact on the observed velocity distribution. In what follows, we therefore consider the properties of the critical binaries with velocity offsets similar to the intrinsic velocity dispersion, and we investigate under which conditions they can be identified as radial velocity variables in multi-epoch data.

\begin{figure*}[htb]
 \centering
 \includegraphics[width=\linewidth]{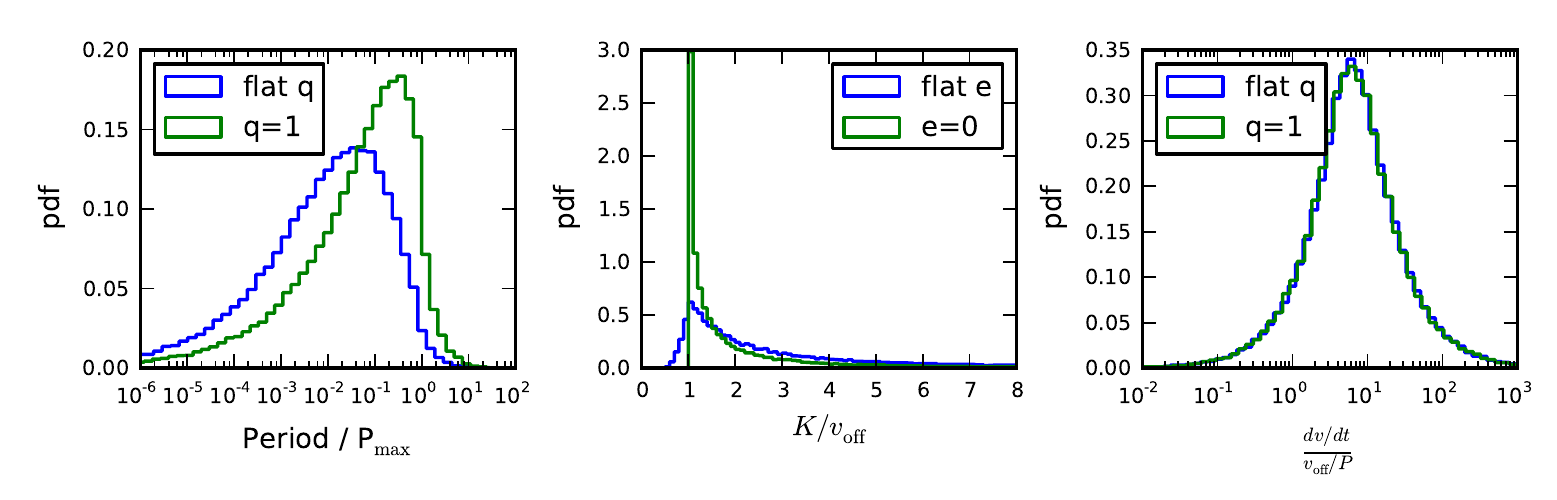}
 \caption{\label{fig:bindist}{\it Left:} the period distribution with respect to the maximum period needed to get a given velocity offset for a circular orbit ($P_{\rm max}$; equation \ref{eq:pmax}) for a flat distribution of mass ratios (blue) or only equal-mass binaries (green). {\it Middle}: the distribution of the radial velocity semi-amplitude ($K$) relative to the velocity offset ($v_{\rm off}$) for a flat distribution of eccentricities and circular orbits. {\it Right:} the distribution of the velocity gradient ($dv/dt$) relative to the velocity offset ($v_{\rm off}$) divided by the orbital period ($P$) for a flat distribution of mass ratios (blue) or only equal-mass binaries (green). These plots are independent of the period distribution of the binaries, and only depend weakly on the mass ratio and eccentricity distributions.} 
 \end{figure*}

For most of the binary orbit, the radial velocity offset is smaller but of the same order of magnitude as the radial velocity semi-amplitude $K$. This is illustrated in the central panel of Figure \ref{fig:bindist}, where we plot the distribution of the radial velocity semi-amplitude over the velocity offset ($K/v_{\rm off}$) at a random point in time for a random set of binaries. The distribution plotted here is independent of the period and mass ratio distribution and only depends on the eccentricity distribution. For both eccentricity distributions considered (flat and only circular), the distribution is strongly peaked at one and decreases quickly with increasing $K/v_{\rm off}$. Only for eccentric orbits can the radial velocity offset be greater than the semi-amplitude. 

If the time between radial velocity observations is sufficiently longer than the period of the binary, the phases of these observations become independent of each other. Radial velocities are drawn from a distribution set only by the radial velocity semi-amplitude $K$ and eccentricity $e$, independent of the orbital period or mass ratio (except for their contribution to $K$). To allow identification as a binary under this long-baseline condition, the measurement uncertainty should be at least smaller than the radial velocity semi-amplitude of the binaries. Thus, to identify the critical binaries with velocity offsets comparable to the intrinsic velocity dispersion, measurement uncertainties lower than the velocity dispersion are required. This is not really an extra requirement, as having measurement uncertainties lower than the velocity dispersion helps to accurately fit the velocity dispersion even when the effect of binaries is ignored. The detection rate of binaries will increase as the measurement uncertainties decrease or the number of epochs increases (by decreasing the probability of the observations being in phase with the binary orbit), however it does not depend on the baseline of the observations, as long as it stays significantly longer than the period of the binary. This explains why, in our Monte Carlo simulations of the systematic offset of the fitted velocity dispersion for two-epoch data (Figure \ref{fig:2epoch}), the systematic velocity offset depends on the measurement uncertainty and not on the baseline once this baseline becomes comparable to the widest binaries in the Monte Carlo simulations (i.e. 300 years).

For any reasonable baseline, there will be many binaries causing radial velocity offsets comparable to the intrinsic velocity dispersion which also have periods much longer than the baseline, in which case the argument above does not hold. To illustrate this, we compute from Kepler's third law the maximum period needed to create a given velocity offset for circular orbits:
\begin{equation}
  \begin{split}
    P_{\rm max} &= P(K=v_{\rm off}, e=0, \sin i = 1, q=1) = \frac{1}{4} \frac{2 \pi G m_1}{v_{\rm off}^3} \\
                    &= 2.6 \times 10^4 \ {\rm yr} \ \frac{m_1}{M_\odot} \left(\frac{v_{\rm off}}{{\rm km\ s}^{-1}}\right)^{-3}. \label{eq:pmax}
   \end{split}
\end{equation}
Projection effects and non-equal mass ratios will result in the typical period of the binary causing a given velocity offset to be lower. We find from randomly sampling a large number of binaries with a flat eccentricity distribution that even for equal-mass binaries the typical period is 0.5 dex smaller (green in left panel of Figure \ref{fig:bindist}). For a flat mass ratio distribution (blue in left panel of Figure~\ref{fig:bindist}), the period distribution of stars causing a given velocity offset is even 1.5 dex smaller than the maximum period given in equation~\ref{eq:pmax}. The distributions plotted in Figure~\ref{fig:bindist} are independent of the period distribution of binaries as they are plotted relative to $P_{\rm max}$. For the R136 data ($\sigma_{\rm v} \sim 6 \ \kms$; $m_1 \sim 30 \ \msun$), $P_{\rm max}$ is about 4\,000 years. Most of the binaries causing a velocity offset comparable to the intrinsic velocity dispersion of the cluster will therefore have periods longer than a baseline of even a few years, although there is a significant minority of binaries with $P < 10^{-3} P_{\rm max} < 4$~years (left panel in Figure \ref{fig:bindist}) for which the period might be shorter than the baseline.

Long-period binaries will be detected when the measurement uncertainty is significantly smaller than the velocity change between the first and last epoch\footnote{Intermediate epochs contribute less to the capability of observing small velocity gradients than the first and last epoch, suggesting that it is better to focus on just two deep epochs than on many less deep epochs with the same total baseline. However, intermediate epochs can still be important in confirming if the velocity differences measured between the first and last epoch are real and they are crucial in solving binary orbits, which allows the binary population to be characterized.}, which we assume below to be separated in time by a baseline $b$. The velocity change is given by $\frac{dv}{dt} b$, where $\frac{dv}{dt}$ is the velocity gradient of the binary, which is roughly constant over the observed time under the condition considered here ($P >> b$). So the detection of a long-period binary depends on whether the velocity gradient is sufficiently larger than the ratio of the measurement uncertainty and baseline ($\sigma_{\rm meas}/b$). In the two-epoch Monte Carlo simulations (Figures \ref{fig:2epoch} and \ref{fig:2epoch_norm}), we indeed find that the systematic offset in the fitted velocity dispersion depends mainly on $\sigma_{\rm meas}/b$ for baselines shorter than a few years. For these baselines, most of the binaries causing velocity offsets comparable to the intrinsic velocity dispersion are in the regime discussed here ($P >> b$). As these binaries need to be detected to reduce the systematic offset, the systematic offset is to first order a function of $\sigma_{\rm meas}/b$ (see Figure \ref{fig:2epoch_norm}), until the baseline becomes comparable to the widest binaries (with periods of $\sim P_{\rm max}$).

To quantify this further, we computed the velocity gradients for a large sample of binaries with flat eccentricity distributions. For binaries causing a given velocity offset $v_{\rm off}$ with orbital period $P$, we find that the velocity gradients follow a roughly log-normal distribution with a peak at $8 v_{\rm off} / P$ and a width of about 0.5 dex (right panel in Figure \ref{fig:bindist}). For R136 ($\sigma_{\rm v} \sim 6 \kms$; $P_{\rm max} \sim 4\,000$~years) detecting the widest binaries causing a velocity offset comparable to the velocity distribution (i.e. with $P \sim P_{\rm max}$) would hence require the detection of a velocity gradient of $8 \sigma_{\rm v} / P_{\rm max} \approx 12$~m~s$^{-1}$~yr$^{-1}$, which is virtually impossible. However, with typical periods being 0.5 - 3 dex lower (left panel in Figure \ref{fig:bindist}), a significant portion of the critical binaries can actually be detected for the typical measurement uncertainty of $1 \ \kms$ and baseline of 1 year. This explains the drop in the systematic offset between the velocity dispersion fitted assuming different binary orbital parameter distributions (which are represented by different colors in Figure \ref{fig:bindist}) when multi-epoch data are used. 

For many lower-mass clusters, the intrinsic velocity dispersion is much lower (often at the sub-$\kms$ level), leading to an increase in the maximum period due to its dependence on the third power of the velocity offset (equation \ref{eq:pmax}). Even though the light of these generally older clusters will be dominated by lower-mass stars (lowering $m_1$ in equation \ref{eq:pmax}), $P_{\rm max}$ is still likely to lie around $10^5-10^7$ years, such that the broadening of the velocity distribution is dominated by binaries with periods of $10^3 - 10^5$ years. These are virtually impossible to detect through spectroscopic means (and will only be detectable in direct imaging surveys for the most nearby clusters). Therefore, although a multi-epoch strategy with a baseline of a few years works well for the large velocity dispersions of young massive clusters, it has little effect on the accuracy with which clusters with a much lower velocity dispersion can be measured, as that would require the detection of binaries with periods of many thousands of years. This is actually fortunate as the need for a multi-epoch strategy is much lower for lower-mass clusters, whose light is dominated by solar-type stars with much better constrained binary properties.

\section{Conclusion \label{conclusion}}

We explored the applicability of the maximum likelihood method presented by \citet{cottaar2012b} to recover the intrinsic velocity dispersion of massive stars in young clusters from a single epoch of radial velocity data. By using Monte Carlo simulations and multi-epoch stellar radial velocity data in the young massive cluster R136 as a test case, we showed that the method works reasonably well, the main limitation being uncertainties in the binary properties of OB stars which can lead to a systematic offset of tens of percent in the fitted velocity dispersion ($< 40\%$ for typical young massive cluster velocity dispersions, i.e. $\gtrsim 4 \ \kms$).

This systematic offset can be greatly reduced by directly detecting the spectroscopic binaries through radial velocity variations in multi-epoch data. Specifically, the spectroscopic binaries with velocity offsets comparable to the intrinsic velocity dispersion need to be detected to increase the accuracy of the fitted velocity dispersion. For a typical young massive cluster, these periods range from $10^{-2}$ to $10^{4}$ years, meaning that many can be detected with typical baselines of observations of a few years. Indeed, we find that the accuracy of a fitted velocity dispersion of $6 \ \kms$ improves to better than 15\% for a typical ratio of the measurement uncertainty over the baseline of $1 \ \kms \ {\rm yr}^{-1}$. However, there will always remain undetected binaries affecting the observed velocity distribution for any reasonable baseline and measurement uncertainty, which will have to be corrected for (see Section~\ref{sec:extension}). As the orbital period scales with the orbital velocity to the third power (Kepler's third law), multi-epoch data are far less useful in improving the accuracy of the fitted velocity dispersion of lower-mass clusters (with sub-$\kms$ velocity dispersions) than of young massive clusters.

Unlike the velocity dispersion, the binary fraction fitted with the method tested in this paper is very sensitive to the assumed binary properties. In general, the method can therefore not reliably recover the binary fraction from a radial velocity dataset if the distributions of orbital parameters of the binaries are not well-constrained.

Although we have tested the method on a sample of radial velocities of O-type stars in R136 and considered binary properties that were derived mainly from spectroscopic surveys of O-type stars\footnote{Early B-type stars were also included in the study of \citet{KK2012}.}, it should also be applicable to samples of B-type stars as they appear to have similar binary properties (Dunstall et al., in preparation). This opens the door to using only one or two epochs of radial velocities to estimate the velocity dispersion of the massive star population in a large number of young Galactic open clusters which may contain at most a few O-type stars but a much larger number of B-type stars.

We may even speculate that this method could eventually be used to study the dynamics of massive stars in systems where this has been almost impossible up to now. There has been an increasing interest in near-infrared spectroscopy of massive stars in recent years \citep[e.g.][]{hanson2005}. This wavelength regime offers the possibility of analyzing these stars in embedded regions and near the Galactic centre due to the substantially lower extinction compared to the optical, and quantitative atmospheric analysis of early-type stars from near-infrared spectral lines is now providing promising results \citep[e.g.][]{stap2011}. Provided that precise radial velocities can be obtained from near-infrared spectra in reasonable exposure times, then the results presented in this paper offer a promising and efficient way of studying the dynamics of massive embedded clusters, or even the dynamics of the significant number of Galactic young massive clusters \citep[e.g.][]{davies2012} which are strongly affected by extinction and for which we currently have very little kinematic information.

\begin{acknowledgements}
We would like to thank the anonymous referee, Hugues Sana, Mark Gieles, Chris Evans, Nate Bastian, and Richard J. Parker for useful feedback on this paper. VHB acknowledges support from the Scottish Universities Physics Alliance (SUPA), the Natural Science and Engineering Research Council of Canada (NSERC), and the ``Fonds the recherche du Qu\'ebec - Nature et technologies" (FRQNT). We also wish to thank the International Space Science Institute (ISSI) in Bern, where this paper was initiated, for supporting Simon Goodwin's international team working on ``The formation of star clusters". \end{acknowledgements}

\bibliographystyle{aa} 
\bibliography{R136_single}

\begin{thebibliography}{55}
\expandafter\ifx\csname natexlab\endcsname\relax\def\natexlab#1{#1}\fi

\bibitem[{{Andersen} {et~al.}(2009){Andersen}, {Zinnecker}, {Moneti},
  {McCaughrean}, {Brandl}, {Brandner}, {Meylan}, \&
  {Hunter}}]{2009ApJ...707.1347A}
{Andersen}, M., {Zinnecker}, H., {Moneti}, A., {et~al.} 2009, \apj, 707, 1347

\bibitem[{{Barb{\'a}} {et~al.}(2010){Barb{\'a}}, {Gamen}, {Arias}, {Morrell},
  {Ma{\'{\i}}z Apell{\'a}niz}, {Alfaro}, {Walborn}, \& {Sota}}]{Barba2010}
{Barb{\'a}}, R.~H., {Gamen}, R., {Arias}, J.~I., {et~al.} 2010, in Revista
  Mexicana de Astronomia y Astrofisica Conference Series, Vol.~38, Revista
  Mexicana de Astronomia y Astrofisica Conference Series, 30--32

\bibitem[{{Bosch} \& {Meza}(2001)}]{boschmeza2001}
{Bosch}, G. \& {Meza}, A. 2001, in Revista Mexicana de Astronomia y Astrofisica
  Conference Series, Vol.~11, Revista Mexicana de Astronomia y Astrofisica
  Conference Series, 29

\bibitem[{{Bosch} {et~al.}(2009){Bosch}, {Terlevich}, \&
  {Terlevich}}]{bosch2009}
{Bosch}, G., {Terlevich}, E., \& {Terlevich}, R. 2009, \aj, 137, 3437

\bibitem[{{Chini} {et~al.}(2012){Chini}, {Hoffmeister}, {Nasseri}, {Stahl}, \&
  {Zinnecker}}]{Chini2012}
{Chini}, R., {Hoffmeister}, V.~H., {Nasseri}, A., {Stahl}, O., \& {Zinnecker},
  H. 2012, \mnras, 424, 1925

\bibitem[{{Clarkson} {et~al.}(2012){Clarkson}, {Ghez}, {Morris}, {Lu},
  {Stolte}, {McCrady}, {Do}, \& {Yelda}}]{clarkson2012}
{Clarkson}, W.~I., {Ghez}, A.~M., {Morris}, M.~R., {et~al.} 2012, \apj, 751,
  132

\bibitem[{{Cottaar} {et~al.}(2012{\natexlab{a}}){Cottaar}, {Meyer}, {Andersen},
  \& {Espinoza}}]{cottaar2012a}
{Cottaar}, M., {Meyer}, M.~R., {Andersen}, M., \& {Espinoza}, P.
  2012{\natexlab{a}}, \aap, 539, A5

\bibitem[{{Cottaar} {et~al.}(2012{\natexlab{b}}){Cottaar}, {Meyer}, \&
  {Parker}}]{cottaar2012b}
{Cottaar}, M., {Meyer}, M.~R., \& {Parker}, R.~J. 2012{\natexlab{b}}, \aap,
  547, A35

\bibitem[{{Crowther} {et~al.}(2010){Crowther}, {Schnurr}, {Hirschi}, {Yusof},
  {Parker}, {Goodwin}, \& {Kassim}}]{Crowther2010}
{Crowther}, P.~A., {Schnurr}, O., {Hirschi}, R., {et~al.} 2010, \mnras, 408,
  731

\bibitem[{{Davies} {et~al.}(2012){Davies}, {de La Fuente}, {Najarro}, {Hinton},
  {Trombley}, {Figer}, \& {Puga}}]{davies2012}
{Davies}, B., {de La Fuente}, D., {Najarro}, F., {et~al.} 2012, \mnras, 419,
  1860

\bibitem[{{De Becker} {et~al.}(2006){De Becker}, {Rauw}, {Manfroid}, \&
  {Eenens}}]{debecker2006}
{De Becker}, M., {Rauw}, G., {Manfroid}, J., \& {Eenens}, P. 2006, \aap, 456,
  1121

\bibitem[{{de Koter} {et~al.}(1998){de Koter}, {Heap}, \& {Hubeny}}]{koterheap}
{de Koter}, A., {Heap}, S.~R., \& {Hubeny}, I. 1998, \apj, 509, 879

\bibitem[{{Elson} {et~al.}(1987){Elson}, {Fall}, \& {Freeman}}]{EFF}
{Elson}, R.~A.~W., {Fall}, S.~M., \& {Freeman}, K.~C. 1987, \apj, 323, 54

\bibitem[{{Evans} {et~al.}(2011){Evans}, {Taylor}, {H{\'e}nault-Brunet},
  {Sana}, {de Koter}, {Sim{\'o}n-D{\'{\i}}az}, {Carraro}, {Bagnoli}, {Bastian},
  {Bestenlehner}, {Bonanos}, {Bressert}, {Brott}, {Campbell}, {Cantiello},
  {Clark}, {Costa}, {Crowther}, {de Mink}, {Doran}, {Dufton}, {Dunstall},
  {Friedrich}, {Garcia}, {Gieles}, {Gr{\"a}fener}, {Herrero}, {Howarth},
  {Izzard}, {Langer}, {Lennon}, {Ma{\'{\i}}z Apell{\'a}niz}, {Markova},
  {Najarro}, {Puls}, {Ramirez}, {Sab{\'{\i}}n-Sanjuli{\'a}n}, {Smartt},
  {Stroud}, {van Loon}, {Vink}, \& {Walborn}}]{VFTS_I}
{Evans}, C.~J., {Taylor}, W.~D., {H{\'e}nault-Brunet}, V., {et~al.} 2011, \aap,
  530, A108

\bibitem[{{Garc{\'{\i}}a} \& {Mermilliod}(2001)}]{garcia2001}
{Garc{\'{\i}}a}, B. \& {Mermilliod}, J.~C. 2001, \aap, 368, 122

\bibitem[{{Garmany} {et~al.}(1980){Garmany}, {Conti}, \&
  {Massey}}]{Garmany1980}
{Garmany}, C.~D., {Conti}, P.~S., \& {Massey}, P. 1980, \apj, 242, 1063

\bibitem[{{Geller} \& {Mathieu}(2011)}]{GellerMathieu2011}
{Geller}, A.~M. \& {Mathieu}, R.~D. 2011, \nat, 478, 356

\bibitem[{{Geller} {et~al.}(2008){Geller}, {Mathieu}, {Harris}, \&
  {McClure}}]{Geller2008}
{Geller}, A.~M., {Mathieu}, R.~D., {Harris}, H.~C., \& {McClure}, R.~D. 2008,
  \aj, 135, 2264

\bibitem[{{Geller} {et~al.}(2009){Geller}, {Mathieu}, {Harris}, \&
  {McClure}}]{Geller2009}
{Geller}, A.~M., {Mathieu}, R.~D., {Harris}, H.~C., \& {McClure}, R.~D. 2009,
  \aj, 137, 3743

\bibitem[{{Gieles} \& {Portegies Zwart}(2011)}]{gielesPZ2011}
{Gieles}, M. \& {Portegies Zwart}, S.~F. 2011, \mnras, 410, L6

\bibitem[{Gieles {et~al.}(2010)Gieles, Sana, \& Portegies~Zwart}]{gieles2010}
Gieles, M., Sana, H., \& Portegies~Zwart, S.~F. 2010, \mnras, 402, 1750

\bibitem[{{Hanson} {et~al.}(2005){Hanson}, {Kudritzki}, {Kenworthy}, {Puls}, \&
  {Tokunaga}}]{hanson2005}
{Hanson}, M.~M., {Kudritzki}, R.-P., {Kenworthy}, M.~A., {Puls}, J., \&
  {Tokunaga}, A.~T. 2005, \apjs, 161, 154

\bibitem[{{H{\'e}nault-Brunet}
  {et~al.}(2012{\natexlab{a}}){H{\'e}nault-Brunet}, {Evans}, {Sana}, {Gieles},
  {Bastian}, {Ma{\'{\i}}z Apell{\'a}niz}, {Markova}, {Taylor}, {Bressert},
  {Crowther}, \& {van Loon}}]{VHB2012}
{H{\'e}nault-Brunet}, V., {Evans}, C.~J., {Sana}, H., {et~al.}
  2012{\natexlab{a}}, \aap, 546, A73

\bibitem[{{H{\'e}nault-Brunet}
  {et~al.}(2012{\natexlab{b}}){H{\'e}nault-Brunet}, {Gieles}, {Evans}, {Sana},
  {Bastian}, {Ma{\'{\i}}z Apell{\'a}niz}, {Taylor}, {Markova}, {Bressert}, {de
  Koter}, \& {van Loon}}]{vhb2012_rot}
{H{\'e}nault-Brunet}, V., {Gieles}, M., {Evans}, C.~J., {et~al.}
  2012{\natexlab{b}}, \aap, 545, L1

\bibitem[{{Hillwig} {et~al.}(2006){Hillwig}, {Gies}, {Bagnuolo}, {Huang},
  {McSwain}, \& {Wingert}}]{hillwig2006}
{Hillwig}, T.~C., {Gies}, D.~R., {Bagnuolo}, Jr., W.~G., {et~al.} 2006, \apj,
  639, 1069

\bibitem[{{Kiminki} \& {Kobulnicky}(2012)}]{KK2012}
{Kiminki}, D.~C. \& {Kobulnicky}, H.~A. 2012, \apj, 751, 4

\bibitem[{{Kleyna} {et~al.}(2002){Kleyna}, {Wilkinson}, {Evans}, {Gilmore}, \&
  {Frayn}}]{Kleyna02}
{Kleyna}, J., {Wilkinson}, M.~I., {Evans}, N.~W., {Gilmore}, G., \& {Frayn}, C.
  2002, \mnras, 330, 792

\bibitem[{{Kobulnicky} \& {Fryer}(2007)}]{KF2007}
{Kobulnicky}, H.~A. \& {Fryer}, C.~L. 2007, \apj, 670, 747

\bibitem[{{Koposov} {et~al.}(2011){Koposov}, {Gilmore}, {Walker}, {Belokurov},
  {Wyn Evans}, {Fellhauer}, {Gieren}, {Geisler}, {Monaco}, {Norris}, {Okamoto},
  {Pe{\~n}arrubia}, {Wilkinson}, {Wyse}, \& {Zucker}}]{koposov2011}
{Koposov}, S.~E., {Gilmore}, G., {Walker}, M.~G., {et~al.} 2011, \apj, 736, 146

\bibitem[{{Lucy}(2006)}]{Lucy2006}
{Lucy}, L.~B. 2006, \aap, 457, 629

\bibitem[{{Mackey} \& {Gilmore}(2003)}]{mackey2003}
{Mackey}, A.~D. \& {Gilmore}, G.~F. 2003, \mnras, 338, 85

\bibitem[{{Mahy} {et~al.}(2009){Mahy}, {Naz{\'e}}, {Rauw}, {Gosset}, {De
  Becker}, {Sana}, \& {Eenens}}]{mahy2009}
{Mahy}, L., {Naz{\'e}}, Y., {Rauw}, G., {et~al.} 2009, \aap, 502, 937

\bibitem[{{Mahy} {et~al.}(2013){Mahy}, {Rauw}, {De Becker}, {Eenens}, \&
  {Flores}}]{mahy2013}
{Mahy}, L., {Rauw}, G., {De Becker}, M., {Eenens}, P., \& {Flores}, C.~A. 2013,
  \aap, 550, A27

\bibitem[{{Ma{\'{\i}}z-Apell{\'a}niz}(2001)}]{maiz2001}
{Ma{\'{\i}}z-Apell{\'a}niz}, J. 2001, \apj, 563, 151

\bibitem[{{Martinez} {et~al.}(2011){Martinez}, {Minor}, {Bullock},
  {Kaplinghat}, {Simon}, \& {Geha}}]{Martinez11}
{Martinez}, G.~D., {Minor}, Q.~E., {Bullock}, J., {et~al.} 2011, \apj, 738, 55

\bibitem[{{Mason} {et~al.}(2009){Mason}, {Hartkopf}, {Gies}, {Henry}, \&
  {Helsel}}]{Mason2009}
{Mason}, B.~D., {Hartkopf}, W.~I., {Gies}, D.~R., {Henry}, T.~J., \& {Helsel},
  J.~W. 2009, \aj, 137, 3358

\bibitem[{{McLaughlin} \& {van der Marel}(2005)}]{mclaughlin2005}
{McLaughlin}, D.~E. \& {van der Marel}, R.~P. 2005, \apjs, 161, 304

\bibitem[{{Odenkirchen} {et~al.}(2002){Odenkirchen}, {Grebel}, {Dehnen}, {Rix},
  \& {Cudworth}}]{Odenkirchen02}
{Odenkirchen}, M., {Grebel}, E.~K., {Dehnen}, W., {Rix}, H.-W., \& {Cudworth},
  K.~M. 2002, \aj, 124, 1497

\bibitem[{{{\"O}pik}(1924)}]{opik1924}
{{\"O}pik}, E. 1924, Publications of the Tartu Astrofizica Observatory, 25, 1

\bibitem[{{Pinsonneault} \& {Stanek}(2006)}]{PS2006}
{Pinsonneault}, M.~H. \& {Stanek}, K.~Z. 2006, \apjl, 639, L67

\bibitem[{{Raghavan} {et~al.}(2010){Raghavan}, {McAlister}, {Henry}, {Latham},
  {Marcy}, {Mason}, {Gies}, {White}, \& {ten Brummelaar}}]{Raghavan10}
{Raghavan}, D., {McAlister}, H.~A., {Henry}, T.~J., {et~al.} 2010, \apjs, 190,
  1

\bibitem[{{Rauw} {et~al.}(2009){Rauw}, {Naz{\'e}}, {Fern{\'a}ndez Laj{\'u}s},
  {Lanotte}, {Solivella}, {Sana}, \& {Gosset}}]{rauw2009}
{Rauw}, G., {Naz{\'e}}, Y., {Fern{\'a}ndez Laj{\'u}s}, E., {et~al.} 2009,
  \mnras, 398, 1582

\bibitem[{{Reggiani} \& {Meyer}(2013)}]{Reggiani13}
{Reggiani}, M. \& {Meyer}, M.~R. 2013, \aap, 553, A124

\bibitem[{{Rochau} {et~al.}(2010){Rochau}, {Brandner}, {Stolte}, {Gennaro},
  {Gouliermis}, {Da Rio}, {Dzyurkevich}, \& {Henning}}]{rochau2010}
{Rochau}, B., {Brandner}, W., {Stolte}, A., {et~al.} 2010, \apjl, 716, L90

\bibitem[{{Sabbi} {et~al.}(2012){Sabbi}, {Lennon}, {Gieles}, {de Mink},
  {Walborn}, {Anderson}, {Bellini}, {Panagia}, {van der Marel}, \& {Ma{\'{\i}}z
  Apell{\'a}niz}}]{sabbi2012}
{Sabbi}, E., {Lennon}, D.~J., {Gieles}, M., {et~al.} 2012, \apjl, 754, L37

\bibitem[{{Sana} {et~al.}(2013){Sana}, {de Koter}, {de Mink}, {Dunstall},
  {Evans}, {H{\'e}nault-Brunet}, {Ma{\'{\i}}z Apell{\'a}niz},
  {Ram{\'{\i}}rez-Agudelo}, {Taylor}, {Walborn}, {Clark}, {Crowther},
  {Herrero}, {Gieles}, {Langer}, {Lennon}, \& {Vink}}]{Sana_2013}
{Sana}, H., {de Koter}, A., {de Mink}, S.~E., {et~al.} 2013, \aap, 550, A107

\bibitem[{{Sana} {et~al.}(2012){Sana}, {de Mink}, {de Koter}, {Langer},
  {Evans}, {Gieles}, {Gosset}, {Izzard}, {Le Bouquin}, \&
  {Schneider}}]{Sana2012_science}
{Sana}, H., {de Mink}, S.~E., {de Koter}, A., {et~al.} 2012, Science, 337, 444

\bibitem[{{Sana} \& {Evans}(2011)}]{SanaEvans2011}
{Sana}, H. \& {Evans}, C.~J. 2011, in IAU Symposium, Vol. 272, IAU Symposium,
  ed. C.~{Neiner}, G.~{Wade}, G.~{Meynet}, \& G.~{Peters}, 474--485

\bibitem[{{Sana} {et~al.}(2009){Sana}, {Gosset}, \& {Evans}}]{sana2009}
{Sana}, H., {Gosset}, E., \& {Evans}, C.~J. 2009, \mnras, 400, 1479

\bibitem[{{Sana} {et~al.}(2008){Sana}, {Gosset}, {Naz{\'e}}, {Rauw}, \&
  {Linder}}]{sana2008}
{Sana}, H., {Gosset}, E., {Naz{\'e}}, Y., {Rauw}, G., \& {Linder}, N. 2008,
  \mnras, 386, 447

\bibitem[{{Sana} {et~al.}(2011){Sana}, {James}, \& {Gosset}}]{sana2011}
{Sana}, H., {James}, G., \& {Gosset}, E. 2011, \mnras, 416, 817

\bibitem[{{Simon} {et~al.}(2011){Simon}, {Geha}, {Minor}, {Martinez}, {Kirby},
  {Bullock}, {Kaplinghat}, {Strigari}, {Willman}, {Choi}, {Tollerud}, \&
  {Wolf}}]{simon2011}
{Simon}, J.~D., {Geha}, M., {Minor}, Q.~E., {et~al.} 2011, \apj, 733, 46

\bibitem[{{Stap} {et~al.}(2011){Stap}, {Sana}, \& {de Koter}}]{stap2011}
{Stap}, F.~A., {Sana}, H., \& {de Koter}, A. 2011, Journal of Physics
  Conference Series, 328, 012025

\bibitem[{{Zahn}(1977)}]{zahn1977}
{Zahn}, J.-P. 1977, \aap, 57, 383

\bibitem[{{Zahn}(1978)}]{zahn1978}
{Zahn}, J.-R. 1978, \aap, 67, 162

\end{thebibliography}

\appendix
\section{Review of the binary properties of OB stars \label{bin_review}}

The current constraints on the binary fraction and distributions of orbital parameters of massive binaries are reviewed below.

\subsection{Period distribution}

\citet{Sana2012_science} showed that the intrinsic period distribution of their Galactic open clusters sample does not follow the widely used \"{O}pik law \citep[i.e., a flat distribution in the logarithm of the separation or, equivalently, of the period; ][]{opik1924} but is instead overabundant in short period systems. They found an exponent $\pi=-0.55\pm0.22$ for the power law of the period distribution for lower and upper bounds of 0.15 and 0.35 on $\log_{10}P/{\rm d}$. For the same period range, \citet{Sana_2013} also found from the 30~Doradus dataset a stronger preference for short periods than previously assumed, with a comparable value of $\pi=-0.45\pm0.30$. This is in contrast with the best-fit value of $\pi=0.2\pm0.4$ from \citet{KK2012} for Cyg~OB2 binaries, which is consistent with the \"{O}pik law to within 1\,$\sigma$, although these authors argued that no single power law adequately reproduces the data at the shortest periods ($P<14$~days). Although this work assumes a power law distribution valid between 1 and 1\,000 days, we must also highlight that the sample does not probe this full range of periods. The power law exponent of the period distribution was determined from 22 binaries having periods shorter than 30 days, so the results had to be extrapolated over almost two orders of magnitude.

\subsection{Mass ratio distribution}

It has been reported that the mass ratio distribution of massive binaries tends to peak towards unity \citep{boschmeza2001, PS2006}, but this has since been contested \citep{Lucy2006} and more recent studies tend to favor a flat distribution of mass ratios. \citet{KK2012} indeed inferred a value of $\kappa=0.1\pm0.5$ for the exponent of the power-law distribution of mass ratios for the known massive binaries in Cyg~OB2. Similarly, \citet{Sana2012_science} found no preference for equal-mass binaries ($\kappa = -0.1\pm0.6$) in Galactic open clusters. 
In 30 Doradus, \citet{Sana_2013} even found a mass ratio distribution that is slightly skewed towards systems with low mass ratios ($\kappa=-1.0\pm0.4$), although this only provides a weak constraint on the distribution of mass ratios, and these results are still in agreement within 2\,$\sigma$ with the two previous studies reporting flat distributions. These results are all incompatible with a random pairing from a classical mass function \citep[i.e. $\kappa=-2.35$; see][]{Sana_2013}.

\subsection{Eccentricity distribution}

Because measuring the eccentricity of a spectroscopic binary requires many epochs of radial velocity data, we are only beginning to probe the eccentricity distribution of massive binaries. As expected from tidal dissipation that tends to circularize their orbit \citep{zahn1977, zahn1978}, a large fraction of the short-period massive binary systems are found to have low eccentricities \citep{SanaEvans2011, KK2012}. \citet{KK2012} found a value of $\eta=-0.6\pm0.3$ for the exponent of the power-law distribution of eccentricities, while \citet{Sana2012_science} obtained $\eta=-0.45\pm0.17$. \citet{Sana_2013} could not constrain $\eta$ in the 30~Doradus study and instead adopted the eccentricity distribution inferred by \citet{Sana2012_science} for the Galactic open clusters sample.

\subsection{Binary fraction}

Number of studies have investigated the observed fraction of spectroscopic binaries among massive stars. \citet{Mason2009} compiled results from the literature to show that 51\% of the Galactic O-type stars investigated by multi-epoch spectroscopy are in fact spectroscopic binaries, while this fraction goes up to 56\% for objects in clusters or OB associations. \citet{Barba2010} obtained a similar fraction, with 60\% of the 240 Galactic O and WN stars in their survey of the southern sky displaying significant radial velocity variations (i.e. $>$10~$\kms$). \citet{Chini2012} also observed a high binary fraction in their spectroscopic survey of Galactic O and B stars in the southern sky. Studies focusing on individual young open clusters or OB associations have reported observed binary fractions between 30 and 60\% \citep[e.g.][]{debecker2006, hillwig2006, sana2008, sana2009, mahy2009, rauw2009, sana2011, mahy2013}, with variations from one cluster to the other compatible with the statistical fluctuations expected given the size of the samples \citep{SanaEvans2011}. Thus, although it has been proposed that the spectroscopic binary fraction might be related to the cluster density \citep[e.g.][]{garcia2001}, the current data is consistent with a common binary fraction in all clusters, at least for O-star rich clusters \citep[for details see][]{SanaEvans2011}.

To constrain the intrinsic fraction of spectroscopic binaries, one has to correct the observed fraction for observational biases which depend on the underlying distributions of orbital parameters. \citet{KF2007} opted to fix the period distribution to the standard \"{O}pik law because the solution of their Monte Carlo simulations for the period and mass-ratio distributions was degenerate. They inferred an intrinsic binary fraction of over 80\%, but note that the range of separations considered in this study (up to 10\,000 AU) extends well beyond the sensitivity domain of their spectroscopic observations. \citet{KK2012} observed a binary fraction of 21\% in Cyg OB2 (24/114 objects) and inferred an intrinsic fraction of $44\pm8\%$ considering binaries with periods between 1 and 1000 days. \citet{Sana2012_science} identified 40 spectroscopic binaries for an observed binary fraction of 56\% in their Galactic open clusters sample and found an intrinsic fraction of $69\pm9\%$ for periods in the range $0.15 < \log_{10}P/{\rm d} < 3.5$. \citet{Sana_2013} observed a spectroscopic binary fraction of $35\pm3\%$ in 30 Doradus, compatible with what \citet{bosch2009} found from a different but overlapping sample of 54 O and early B-type stars, and inferred an intrinsic binary fraction of $51\pm4\%$ for periods in the range $0.15 < \log_{10}P/{\rm d} < 3.5$. This binary fraction appears mostly uniform across the 30 Doradus region and independent of the spectral subtype and luminosity class. 
\end{document}